\documentclass[pdflatex,sn-mathphys-num]{sn-jnl}

\usepackage{graphicx}%
\usepackage{multirow}%
\usepackage{amsmath,amssymb,amsfonts}%
\usepackage{amsthm}%
\usepackage{mathrsfs}%
\usepackage[title]{appendix}%
\usepackage{xcolor}%
\usepackage{textcomp}%
\usepackage{manyfoot}%
\usepackage{booktabs}%
\usepackage{algorithm}%
\usepackage{algorithmicx}%
\usepackage{algpseudocode}%
\usepackage{listings}%
\usepackage{bm}
\usepackage{natbib}
\usepackage{longtable}

\usepackage{array}
\usepackage[export]{adjustbox}
\usepackage{rotating}
\usepackage{hyperref}
\usepackage{caption}
\usepackage{subcaption}
\usepackage{multirow}
\usepackage{booktabs}
\usepackage{bbm}
\usepackage{geometry}

\usepackage{lineno}

\begin{document}

\title[A latent Gaussian CAR copula model for analyzing trends in rainfall]{A Bayesian latent Gaussian conditional autoregressive copula model for analyzing spatially-varying trends in rainfall}

\author[]{\fnm{Sayan} \sur{Bhowmik}}\email{sayanb22@iitk.ac.in}

\author*[]{\fnm{Arnab} \sur{Hazra}}\email{ahazra@iitk.ac.in}



\affil[]{\orgdiv{Department of Mathematics and Statistics}, \orgname{Indian Institute of Technology Kanpur}, \orgaddress{\city{Kanpur}, \postcode{208016}, \country{India}}}



\abstract{Assessing the availability of rainfall water plays a crucial role in rainfed agriculture. Given the substantial proportion of agricultural practices in India being rainfed and considering the potential trends in rainfall amounts across years due to climate change, we build a statistical model for analyzing monsoon total rainfall data for 34 meteorological subdivisions of mainland India available for 1951--2014. Here, we model the marginal distributions using a gamma regression model and the dependence through a Gaussian conditional autoregressive (CAR) copula model. Due to the natural variation in the monsoon total rainfall received across various dry through wet regions of the country, we allow the parameters of the marginal distributions to be spatially varying, under a latent Gaussian model framework. The neighborhood structure of the regions determines the dependence structure of both the likelihood and the prior layers, where we explore both CAR and intrinsic CAR structures for the priors. The proposed methodology also effectively imputes the missing data. We use the Markov chain Monte Carlo algorithms to draw Bayesian inferences. In simulation studies, the proposed model outperforms several competitors that do not allow a dependence structure at the data or prior layers. Implementing the proposed method for the Indian areal rainfall dataset, we draw inferences about the model parameters and discuss the potential effect of climate change on rainfall across India. While the assessment of the impact of climate change on rainfall motivates our study, the proposed methodology can be easily adapted to other contexts dealing with non-Gaussian non-stationary areal datasets where data from single or multiple temporal covariates are also available, and it is appropriate to assume their coefficients to be spatially varying.}




\maketitle

\section{Introduction}
\label{sec:intro}

Rainfall is a crucial climatic factor that influences various sectors, including agriculture, hydrology, and disaster management. It is essential to analyze and model the spatial and temporal variations in rainfall, especially in monsoon-driven regions like India, where more than 60\% of its agricultural land relies on rainfed farming, making India the most considerable such region globally \citep{venkateswarlu2011rainfed, hazra2024robust}. Climate change has intensified these uncertainties in recent decades, highlighting the need for statistical models that effectively capture rainfall patterns and dependencies \citep{allan2023intergovernmental}. Climate change alters weather patterns, and these shifts influence the distribution of temperature, wind, and rainfall, leading to variations in spatial rainfall patterns. While certain regions may experience increased and more frequent rainfall, others may face droughts and declining precipitation. With varying marginal behavior across different global regions, rainfall exhibits inherent spatial dependence, shaped by large-scale atmospheric circulation, geographical features, and local climatic conditions \citep{guhathakurta2008trends, hazra2024efficient}. Hence, these phenomena motivate us to study the varying trends in rainfall across the meteorological subdivisions of India, while allowing for a realistic spatial dependence structure exhibited by the data.

Historically, the statistical modeling of monthly, seasonal, or annual total rainfall has remained an important research area in meteorology. Monsoon total rainfall data are usually nonzero, and the histograms appear positively skewed; for all meteorological subdivisions of India, the recorded monsoon total rainfall data do not include any year with entirely nil rainfall. Hence, the justified probability distributions here are right-skewed and supported over the positive real line; illustrative examples are exponential \citep{todorovic1975stochastic, hazra2018bayesian}, gamma \citep{barger1949evaluation, mooley1968application}, log-normal \citep{kwaku2007characterization, mandal2015estimation}, Weibull \citep{duan1995comparison, lana2017rainfall}, and generalized exponential \citep{madi2007bayesian, hazra2025minimum} distributions. However, meteorologists have historically favored the gamma distribution for modeling rainfall data \citep{martinez2019precipitation}. While the data across years are often assumed to be distributed identically in the meteorology literature, it is more appropriate to consider a nonstationary behavior across years in the marginal distribution due to the potential trends in meteorological variables driven by climate change. For this purpose, a gamma regression model is a possible approach that various researchers have explored for different scientific disciplines \citep{nelder1972generalized, amin2020performance}.

While the gamma regression can effectively model the nonstationary marginal behavior in rainfall, a fundamental limitation of many traditional univariate models is their failure to incorporate spatial dependency, which is critical for accurate estimation and prediction \citep{diggle1998model}. Copulas \citep{sklar1959distribution, nelsen2006introduction, chen2019copulas} are powerful statistical tools that enable the modeling of complex dependencies between multiple random variables, even extending beyond traditional linear correlation measures \citep{frees1998understanding, omidi2022spatial}. In spatial statistics, copulas facilitate the construction of joint distributions by separately modeling marginal distributions and their spatial dependence structure \citep{sang2010continuous, huser2016non}; this separation allows for greater flexibility in capturing nonlinear and asymmetric dependencies that are often present in spatial data, leading to improved inference and prediction in various applications \citep{graler2011pair,okhrin2017copulae, nazeri2022multivariate, nazeri2023application}. \cite{zhang2012bivariate} attempt to combine the copula theory with the entropy theory for bivariate rainfall and runoff analysis. Further applications of copulas, e.g., Archimedean copulas for meteorological data, are covered in \cite{najjari2012application, zhang2007bivariate}. \cite{mesbahzadeh2019joint} predict temporal trends of precipitation and temperature using a copula under a climate change scenario. In our context of areal data, \cite{musgrove2016hierarchical} propose a copula-based hierarchical model with covariance selection for unbiased estimation of marginal parameters, providing a dependence structure with intuitive conditional and marginal interpretations. They develop a computational framework that permits efficient frequentist inference for their model, even for large datasets. There has been substantial work on copula models with covariate-dependent marginal distributions and dependence structures in recent years. These approaches extend beyond classical static copulas by allowing the copula or marginal parameters to vary with explanatory variables. For example, \cite{patton2006modelling} introduce conditional copulas for time-series applications, while \cite{acar2011dependence} and \cite{klein2016simultaneous} develop regression-type frameworks where covariates enter the copula parameter. Similarly, \cite{abegaz2012semiparametric} and \cite{krupskii2018factor} proposed models with covariate-dependent marginal distributions combined with flexible copula families. Recent contributions include covariate-adjusted vine copulas and spatially varying copula models \citep{czado2019analyzing}. This surge of literature demonstrates a growing interest in models that jointly accommodate marginal heterogeneity and dependence driven by observed or latent covariates.

Several studies have demonstrated the efficacy of using gamma marginal distributions combined with copula models to analyze rainfall patterns. For instance, \cite{khan2020novel} employ gamma marginal distributions within different copula models to capture spatial dependencies in precipitation data. On the other hand, \cite{lee2018multisite} introduces a simulation-based method to enhance the preservation of cross-correlations in multisite precipitation simulations, addressing limitations observed in direct and indirect estimation methods. Here, the author explores a Gaussian copula with identical gamma marginal distributions for data collected at twelve nearby spatial locations; however, the model assumptions are unsuitable for a large geographical domain. Similarly, \cite{van2011copula} utilize copulas with gamma marginal distributions to model the relationship between coarse- and fine-scale rainfall depths. The study in \cite{baxevani2015spatiotemporal} describes precipitation intensity using space and time-dependent gamma distributions. While most of these studies do not assume a gamma regression framework for the marginal distributions, the rest do not focus on annual trends. Besides, the copulas used in these papers are for continuous spatial processes; hence, they are not directly applicable to areal data modeling.

Outside the avenue of copula-based modeling, the two most popular frameworks for areal data modeling are conditional autoregressive (CAR) models \citep{besag1974spatial} and simultaneous autoregressive (SAR) models \cite{cliff1981sar}. While SAR models are used widely in econometrics, regional science, and social sciences, CAR models are popular in Bayesian spatial statistics, epidemiology, and environmental studies. Among numerous uses of the CAR model in the environmental statistics literature, \cite{banerjee2015hierarchical} use hierarchical CAR models for temperature variability analysis, \cite{villarini2010flood} analyze flood risks and rainfall patterns using CAR models, and \cite{cressie2011statistics} use integrated CAR models with spatial dynamic processes for environmental prediction. In most of these models, the data are assumed to be Gaussian, conditioning on the parameters and the latent variables. However, using the CAR model as a copula in the context of environmental statistics or other scientific disciplines is not well-known. Several works have extended areal models beyond the Gaussian framework to accommodate non-Gaussian outcomes such as counts, rates, rainfall indices, or other skewed environmental data. \cite{besag1974spatial} introduce CAR models, which specify the distribution of a spatial variable in a given region conditional on the values in neighboring regions, laying the foundation for modeling spatial dependence in areal data. Subsequent Bayesian hierarchical extensions of CAR models have been developed for disease mapping \citep{best2005comparison, lee2011comparison, jin2005generalized}, allowing more flexible modeling of spatially structured risk. More general treatments of hierarchical non-Gaussian areal models can be found in \cite{banerjee2015hierarchical}. Beyond disease mapping, ecology, demography, and environmental sciences applications have used Poisson or Binomial CAR-type models to handle count and rate data \citep{wakefield2007disease, banerjee2015hierarchical}. In recent years, researchers have also developed flexible extensions for overdispersed or skewed data, such as negative binomial CAR models \cite{lawson2018bayesian}, zero-inflated areal models \cite{neelon2010bayesian}, and generalized additive mixed models with spatial random effects for non-Gaussian responses \cite{wood2017generalized}. Against this backdrop, our contribution differs in two key respects. First, we focus on gamma regression-based marginal distributions tailored to monsoon rainfall totals, linking parameters directly to spatial regions. Second, we combine the marginal distributions with a Gaussian CAR copula that captures spatial dependence across regions. To our knowledge, this combination is novel and has not been applied in the context of large-scale rainfall analysis.

Given the large spatial extent of the meteorological subdivisions of India, with several subdivisions in the western parts of the country receiving very low rainfall and specific subdivisions in the east covering the wettest locations of the globe, the requirement of spatially varying coefficients (SVCs) in gamma regression is obvious \citep{gelfand2003spatial}. Additionally, the regression coefficients are likely to express local homogeneity; hence, we can model the priors for the SVCs within the framework of Gaussian graphical models \citep{banerjee2015hierarchical}. The CAR and intrinsic CAR \citep[ICAR,][]{mollie1996bayesian} priors are popular choices for SVCs. While \cite{cressie2011statistics} discuss ICAR priors in spatio-temporal ecological models, \cite{banerjee2015hierarchical} use ICAR priors for modeling spatial rainfall variability. \cite{fernandez2010bayesian}, \cite{lee2015bayesian}, and many others use the CAR priors, while the \texttt{R} package \texttt{CARBayes} \citep{lee2017spatio} implements various Bayesian areal data models with CAR priors.

In this paper, we develop a statistical model to analyze monsoon total rainfall data for 34 meteorological subdivisions of mainland India from 1951 to 2014. Given the suitability of the gamma distribution for modeling rainfall data and the possible trend in the marginal distribution due to climate change, we assume a gamma regression model for the marginal distributions, where we consider a logarithmic link for the mean structure, written as a linear function of the years (suitably centered and scaled). We consider a Gaussian CAR copula for modeling the dependence structure, where the adjacency is determined by whether or not two meteorological subdivisions share a boundary. While the usual CAR models exhibit nonstationary marginal distributions due to boundary effects, we appropriately scale the CAR covariance matrix to ensure that the copula conditions are satisfied. We allow varying gamma regression coefficients within a latent Gaussian model framework to account for the natural variation in total rainfall during the monsoon season across different climatic regions. Similar to the likelihood layer, the neighborhood structure of the regions determines the dependence structure in prior layers, where we explore CAR and ICAR priors. Our methodology also effectively imputes missing data by assuming Missing at Random (MAR) \citep{Rubin1976}. We draw Bayesian inference using Markov chain Monte Carlo algorithms, specifically, using a Metropolis-within-Gibbs sampler. To quantify the importance of modeling spatial dependence at the likelihood and prior levels, we compare four model variants: (i) independent likelihood and independent priors, (ii) independent likelihood and CAR/ICAR priors, (iii) CAR-copula likelihood and independent priors, and (iv) CAR-copula likelihood and CAR/ICAR priors. Using various criteria for model comparison, we then compare the performances of the proposed and competing models in terms of model fitting and estimation uncertainties. Applying this method to the Indian areal rainfall dataset, we draw inferences about the model parameters and examine the potential impacts of climate change on rainfall patterns across India. The primary objective of this work is to focus on a specific scientific problem, build a statistical model tailored to that problem, validate the approach through simulation studies, and draw inferences based on the fitted model. In doing so, the contribution does not fall solely within the realm of pure statistical methodology, since we do not aim to provide a fully general modeling solution for areal datasets with spatially varying coefficients; nor is it purely interdisciplinary, as we do not restrict ourselves to applying existing tools without methodological input. Instead, the work lies in between, motivated by an important applied context, while simultaneously contributing to modeling ideas of interest to the statistical community. Specifically, we formulate a Bayesian latent Gaussian model that integrates spatially varying gamma regression with a Gaussian CAR copula, enabling us to model skewed distributions, spatial dependence, and region-specific variability in rainfall. While gamma regression and CAR priors have each been used separately in environmental statistics, to our knowledge, their combination within a copula framework for areal rainfall data has not been previously explored. This contribution is particularly relevant for statisticians seeking flexible models for non-Gaussian, spatially dependent areal data beyond the specific rainfall application. The Indian rainfall dataset serves as a motivating case study to demonstrate the practical utility of the proposed method. Still, the modeling framework applies to various spatial datasets in climate science, epidemiology, and environmental monitoring.

We structure the remainder of this article as follows. Section \ref{sec:background} provides a brief background on gamma regression, conditional autoregressive models, and copulas. In Section \ref{sec:eda}, we describe the details of the Indian rainfall dataset we analyze and provide an exploratory analysis motivating us to choose specific components in the proposed statistical methodology. Section \ref{sec:methodology} details the construction of the conditional autoregressive copula for modeling spatial dependence, a gamma regression with SVCs for marginal distributions, and the CAR and ICAR prior structures for the coefficients. In Section \ref{sec:computation}, we discuss an MCMC sampling strategy employed for parameter estimation. Section \ref{sec:simulation} discusses a simulation study to validate the correctness of the proposed parameter estimation procedure and to compare the proposed model with simpler alternatives that ignore spatial dependence structures. In Section \ref{sec:application}, we apply the proposed and competing methods to the Indian rainfall dataset and discuss the results. Finally, Section \ref{sec:conclusion} concludes with a discussion on key findings, implications for climate studies, and future research directions.

\section{Background}
\label{sec:background}

This section provides a brief overview of the key statistical tools required for our methodology and data analysis, including gamma regression for handling nonstationary, skewed, and positive-valued data, conditional autoregressive (CAR) models for areal datasets, and copulas. We also discuss two exploratory tools: the Quantile-Quantile (QQ) plot of uniform-transformed data for marginal model validation and Moran's I for assessing spatial autocorrelation. Finally, we mention the Deviance Information Criterion (DIC) and the Watanabe-Akaike Information Criterion (WAIC) for Bayesian model comparison.

\subsection{Gamma regression}
\label{subsec:gamma_reg}

In a gamma regression model \citep{nelder1972generalized, McCullagh2019glm} for $n$ independent but non-identically distributed positive-valued observations, we assume that the response variable $Y_{i}$ follows 
\begin{equation}\label{eq:gamma_reg_definition}
    Y_i \overset{\textrm{Indep}}{\sim} \text{Gamma}(\mu_i, a),~~i=1,\ldots, n,
\end{equation}
where $\mu_i$ and $a$ represent the mean and shape parameter, respectively. The density of the gamma distribution in \eqref{eq:gamma_reg_definition} is given by
$$f(y_i) = \frac{(a / \mu_i)^a}{\Gamma(a)} y_i^{a-1} \exp[-ay_i/\mu_i], ~~ y_i > 0, \mu_i >0, a>0.$$
The relationship between $\mu_i$ and covariates $\bm{x}_i$ is modeled through a log-link function
\begin{equation}
    \log(\mu_i) = \bm{x}_i^\intercal \bm{\beta},~~i=1,\ldots, n,
\end{equation}
where $\bm{\beta}$ denotes the regression coefficients. This framework makes the marginal variance $\text{Var}(Y_i) = \mu_i^2/a \propto \mu_i^2$. Here, the rate parameter can be calculated as
$$\lambda_{i} = a / \mu_{i} = a~ \exp[-\bm{x}_i^\intercal \bm{\beta}] = \exp[\log(a) - \bm{x}_i^\intercal \bm{\beta}].$$
Here, the coefficient of variation $\sqrt{\text{Var}(Y_i)} / \text{E}(Y_i) = a^{-1/2}$ does not depend on $\mu_i$. This property allows us to interpret the shape parameter $a$ as controlling relative dispersion independently of the mean, which will later help us distinguish between regions with similar average rainfall but different variability.

\subsection{Conditional Autoregressive (CAR) models}
\label{subsec:car_models}

CAR models specify a spatial dependence structure where the conditional distribution of the response variable at each region, given all other areas, depends only on the responses of the neighboring areas \citep{besag1974spatial, cressie1993statistics}. Formally, let $Y_i$ denote the response at region $i$, and let $i \sim j$ indicate regions $i$ and $j$ are neighbors. The neighborhood structure is encoded in a spatial weight matrix $\bm{W}$, with its $(i, j)$th element given by
\begin{equation}\label{eq:neighborhood_weight}
    w_{ij} = \begin{cases}
        1 & \text{if } i \sim j, \\
        0 & \text{if } i \nsim j,
    \end{cases}
\end{equation}
and $w_{ii}=0$ for all $i$. The number of neighbors of region $i$ is $m_i = \sum_{j} w_{ij}$, and we denote by $\bm{M}$ the diagonal matrix with entries $m_{ii}=m_i$.

Given a set of regionally indexed observations, the CAR model is defined through the set of full conditional distributions
\begin{equation} \label{eq:car_model}
    Y_{i}\,\big|\,\{Y_j : j \neq i\} \;\sim\; \textrm{Normal}\!\left( 
    \rho \sum_{j \in \{i : i \sim j, i\ne j\}} \frac{w_{ij}}{m_i}\, Y_{j},\; \frac{\sigma^2}{m_i} 
    \right),
\end{equation}
where $\rho$ is a spatial autocorrelation parameter that controls the strength of dependence among the regions, and $\sigma^2$ is a marginal variance parameter that controls the overall spatial covariance. Here~\eqref{eq:car_model} shows that the conditional mean of $Y_i$ is a weighted average of the values of its neighbors, and the conditional variance decreases with the number of neighbors.

By Brook's lemma \citep{brook1964distinction}, the above conditionals correspond to a zero-mean Gaussian Markov random field (GMRF) with precision matrix
\[
    \bm{Q} = \frac{1}{\sigma^2}\, (\bm{M} - \rho \bm{W}),
\]
so that the covariance matrix is $\sigma^2 (\bm{M} - \rho \bm{W})^{-1}$ whenever this inverse exists. Because the conditional expectations in \eqref{eq:car_model} are expressed entirely in terms of neighboring observations, without any additive constants, the implied joint distribution has a zero mean vector \citep{besag1974spatial, rue2005gaussian, banerjee2015hierarchical}.

A proper CAR model arises when the precision matrix $\bm{Q}$ is positive definite, i.e., when $|\rho|<1$ under standard neighborhood choices; this ensures that the joint distribution is a valid multivariate normal distribution. In contrast, when $\rho=1$, the precision matrix is singular, leading to the intrinsic CAR (ICAR) model. The ICAR model does not define a proper joint distribution. Still, it is widely used as an improper prior for latent spatial random effects (e.g., spatially varying coefficients). CAR models thus provide a convenient mechanism for local spatial smoothing, borrowing strength across neighboring areas, and improving predictive accuracy and inference.

\subsection{Copulas}
\label{subsec:copulas}

Copulas offer a flexible approach to modeling and estimating the joint distribution of variables, independent of their marginal distributions, making them useful for capturing complex dependencies \citep{sklar1959distribution, nelsen2006introduction}. A copula is a joint probability distribution function that ensures both marginal distributions are $\textrm{Uniform}(0, 1)$. Specifically, a function $C(y_1, y_2)$ qualifies as a copula if $C(0, 0) = 0$ and for $y_1, y_2 \in (0, 1)$, it satisfies $C(y_1, 1) = y_1,\ C(1, y_2) = y_2.$ Suppose we aim to determine a suitable joint probability distribution function $H(y_1, y_2)$ for random variables $Y_1$ and $Y_2$, given that their marginal distributions follow the continuous distribution functions $F_1$ and $F_2$, respectively. That is, given $F_1(y_1)$ and $F_2(y_2)$, along with knowledge of the dependency structure between $Y_1$ and $Y_2$, we seek to define an appropriate joint distribution function as $H(y_1, y_2) = P(Y_1 \leq y_1, Y_2 \leq y_2)$. Since the transformed variables $U_1 = F_1(Y_1)$ and $U_2 = F_2(Y_2)$ follow a uniform distribution on $(0, 1)$, the joint distribution function of $U_1$ and $U_2$ forms a copula. Then
\begin{equation} \label{eq:copula}
    \begin{split}
        H(y_1, y_2) & = P(Y_1 \leq y_1, Y_2 \le y_2) \\ & = P(F_1(Y_1) \leq F_1(y_1), F_2(Y_2) \le F_2(y_2)) = C(F_1(y_1), F_2(y_2)).
    \end{split}
\end{equation}
In our context, we consider a random vector following a standard bivariate Gaussian distribution with correlation $\rho$. The Gaussian copula function is defined as $C_\rho(u, v) = \Phi_\rho(\Phi^{-1}(u), \Phi^{-1}(v))$, where $u, v \in [0, 1], \Phi(\cdot)$ represents the standard normal cumulative distribution function (CDF), and $\Phi_\rho(\cdot, \cdot)$ denotes the CDF of a standard bivariate Gaussian distribution with correlation $\rho$. To construct a bivariate Gamma distribution, we assume that the random vector $(Y_1, Y_2)$ has gamma marginal distributions given by the distribution functions $F_1$ and $F_2$, respectively. Using the Gaussian copula, we define  $(Y_1, Y_2) = \left(F_1^{-1}(\Phi(X_1)), F_2^{-1}(\Phi(X_2))\right)$, where $F_1^{-1}$ and $F_1^{-1}$ are the marginal quantile functions for $Y_1$ and $Y_2$, and $(X_1, X_2)\sim \Phi_\rho$. The joint CDF of $Y_1$ and $Y_2$ is then given by $H(Y_1, Y_2; \rho) = C_\rho (\Phi(X_1), \Phi(X_2)),$ while ensuring that the marginal distributions of $Y_1$ and $Y_2$ remain $F_1$ and $F_2$, respectively.

An appropriate copula should accurately represent the assumed dependencies between $U_1$ and $U_2$. Since $F_1$ and $F_2$ are increasing functions, the dependence structure induced by the chosen copula should reflect the relationship we expect between $Y_1$ and $Y_2$. For example, if we assume that $Y_1$ and $Y_2$ have a correlation of $\rho$, we should select a copula such that the random variables following this copula exhibit a correlation of $\rho$. However, since correlation captures only linear dependencies, the correlation between $Y_1$ and $Y_2$ does not necessarily match the correlation between $U_1$ and $U_2$ \citep{ross2022simulation}.

While the above exposition uses two random variables for clarity, copulas extend naturally to higher dimensions. For a random vector $(Y_1, Y_2, \dots, Y_n)^\intercal$ with continuous marginal distributions $F_1, F_2,\dots, F_n$, Sklar’s theorem \citep{sklar1959distribution} suggests the existence of a unique copula $C$ such that for joint CDF $H$ of $(Y_1, Y_2, \dots, Y_n)$, we have 
\begin{equation} \label{eq:multivariate_copula}
    \begin{split}
        H(y_1, y_2, \dots, y_n) & = P(Y_1 \leq y_1, Y_2 \le y_2, \dots, Y_n \le y_n) \\
        & = P(F_1(Y_1) \leq F_1(y_1), F_2(Y_2) \le F_2(y_2), \dots, F_n(Y_n) \le F_n(y_n)) \\
        & = C(F_1(y_1), F_2(y_2), \dots, F_n(y_n)).
    \end{split}
\end{equation}
While the above generalization for any dimension $n$ is possible, only a few classes of copulas, like the Gaussian or the Student-$t$ copula, can be easily computed when $n$ is large. In our application, we employ a Gaussian copula, where we parameterize the dependence through the correlation (or precision) matrix that reflects the neighborhood structure of the Indian meteorological subdivisions.

\subsection{Uniform-transformed Quantile-Quantile (QQ) plot}
\label{subsec:uniform_qq}

Model validation is essential to ensure the reliability of statistical inference. Suppose we aim to fit a marginal distribution $F(\cdot, \bm{\theta}_i)$ for each observation $Y_i$, where the parameter $\bm{\theta}_i$ may vary across $i$. One can often explain such variation through covariates under a regression framework. Based on estimates of $\bm{\theta}_i$, say $\widehat{\bm{\theta}}_i$, approximately $Y_i \sim F(\cdot, \widehat{\bm{\theta}}_i)$ and thus by the probability integral transform, $F(Y_i, \widehat{\bm{\theta}}_i) \sim \textrm{Uniform}(0,1)$ if the model $F$ is appropriate, meaning that it correctly represents the true marginal distribution of $Y_i$. We can thus consider a QQ plot of low through high quantiles of the $\textrm{Uniform}(0,1)$ distribution and the corresponding quantiles of the empirical CDF of $F(Y_i, \widehat{\bm{\theta}}_i)$s as a diagnostic tool. We refer to this tool as a Uniform-transformed QQ plot \citep{hazra2020multivariate}. If the plot remains close to the $y=x$ line, we may conclude that $F$ provides a reasonable fit, in the sense that the model-implied distributions are consistent with the observed data. Conversely, systematic deviations highlight model misspecification, for example, a good fit in the bulk but a poor fit in the tails. If we compare two models, $F_1$ and $F_2$, we can calculate and compare the uniform-transformed QQ plots for both distributions.

Beyond visual inspection, one can quantify the departure from the $y=x$ line to yield a numerical goodness-of-fit measure. Two simple options are the root mean squared error and the mean absolute error, given by
\[
    \text{RMSE}_{U} = \sqrt{\frac{1}{n} \sum_{i=1}^n \left( U_{(i)} - \frac{i}{n+1} \right)^2},~~~~\text{MAE}_{U} = \frac{1}{n} \sum_{i=1}^n \left\vert U_{(i)} - \frac{i}{n+1} \right\vert,
\]
respectively, where $U_{(i)}$ denotes the $i$th order statistic of $\{U_i\}$. This statistic measures the average squared distance between the empirical and theoretical quantiles, directly corresponding to deviations in the QQ plot. More formal alternatives include classical goodness-of-fit statistics such as the Kolmogorov–Smirnov \citep{Kolmogorov1933, smirnov1948table}, Cramér–von Mises \citep{d1986goodness}, or Anderson–Darling tests \citep{anderson1954test}, each emphasizing different aspects of deviation (maximum discrepancy, average discrepancy, or tail behavior). Thus, the uniform-transformed QQ plot provides a graphical diagnostic, and one can pair it with a scalar summary for model comparison. We will use this diagnostic tool later in Section \ref{sec:eda} to validate the choice of gamma marginal distributions compared to alternatives such as the log-normal distribution.

\subsection{Moran’s I for assessing spatial autocorrelation}
\label{subsec:morans_I}

Areal data often exhibit autocorrelation, meaning that observations at nearby, possibly neighboring, regions are more similar to each other than to those at distant ones. Moran’s I \citep{moran1950notes} is a widely used measure for detecting spatial autocorrelation, and it is calculated as
\begin{equation}\label{eq:morans_I}
    I = \frac{N}{\sum_{i}\sum_j w_{ij}} \times \frac{\sum_{i}\sum_{j}w_{ij}(Y_{i} - \bar{Y})(Y_{j} - \bar{Y})}{\sum_{i}(Y_{i} - \bar{Y})^2},
\end{equation}
where $Y_{i}$ represents the observed value at region $i$, $w_{ij}$ is the spatial weight matrix defining neighborhood relationships in \eqref{eq:neighborhood_weight}, $\bar{Y}$ is the mean value of $Y_{i}$, and $N$ is the total number of locations. Moran’s I ranges from $-1$ to $+1$, where positive values of Moran’s I suggest clustering, negative values indicate dispersion, and values near zero suggest spatial randomness. Under the null hypothesis of complete randomness, $\text{E}(I) = -1/(N-1)$ and $\text{Var}(I)$ can be calculated in terms of $w_{ij}$s, and we thus can obtain a $z$-score by centering and scaling. Finally, we obtain the $p$-value by assuming normality for the test statistic and decide on the spatial randomness accordingly.

\subsection{Bayesian model comparison}
\label{subsec:dic_waic}

To compare multiple Bayesian models with the same likelihood structure but different prior structures, we report Deviance information criteria \citep[DIC,][]{spiegelhalter2002bayesian} and Watanabe-Akaike information criteria \citep[WAIC,][]{watanabe2010asymptotic}. We calculate DIC as
\begin{equation*}
    \textrm{DIC}=\bar{D}+p_D=D(\bm{Y} \mid \widehat{\bm{\theta}})+2 p_D,
\end{equation*}
where $D(\bm{Y} \vert \bm{\theta}) = -2 \log[{f}(\bm{Y} \vert \bm{\theta})]$ is the deviance for measuring model fit, $\bar{D}$ is the posterior mean deviance defined as $\text{E}(D(\bm{Y} \mid \bm{\theta}) | \bm{Y})$, $\widehat{\bm{\theta}} = \text{E}(\bm{\theta} | \bm{Y})$ is the posterior mean of the model parameter vector $\bm{\theta}$, and $p_D = \bar{D} - D(\bm{Y} \mid \widehat{\bm{\theta}})$ is the effective number of parameters in the model. Given the MCMC outputs, calculating DIC is straightforward. The intuition is that models with smaller DIC are parsimonious (small $p_D$) and fit well (small $\bar{D}$).

Further, we can compute WAIC, the widely applicable information criterion, as an alternative to DIC. Here, WAIC approximates $n$-fold (i.e., leave-one-out) cross-validation. Suppose we have $n$ independent observations (possibly vectors) $\bm{Y}_i$ for $i=1,\ldots, n$ and $\bm{\theta}$ is the parameter vector in the model (likelihood) $f(\cdot)$. Let $m_i$ and $v_i$ be the posterior mean and variance of $\log \left[f\left(\bm{Y}_{\bm{i}} \mid \bm{\theta}\right)\right]$. The effective model size is $p_W=\sum_{i=1}^n v_i$ and WAIC is given by
\begin{equation*}
    \textrm{WAIC}= -2 \sum_{i=1}^n m_i+2 p_W.
\end{equation*}
WAIC estimates complexity based on the variance of the log-likelihood across posterior samples; this accounts for models with hierarchical structures or non-Gaussian posteriors where the parameter count is ambiguous.

The measure DIC is computationally simpler than WAIC but may be unreliable for hierarchical and complex models. At the same time, WAIC is more general and provides a better approximation of predictive performance, making it preferable in most Bayesian applications. \citep{vehtari2017practical} recommend using WAIC or leave-one-out cross-validation over DIC for model comparison in modern Bayesian workflows.

\section{Data description and exploratory analysis}
\label{sec:eda}
\begin{figure}[t]
    \centering
    \includegraphics[width=0.32\linewidth]{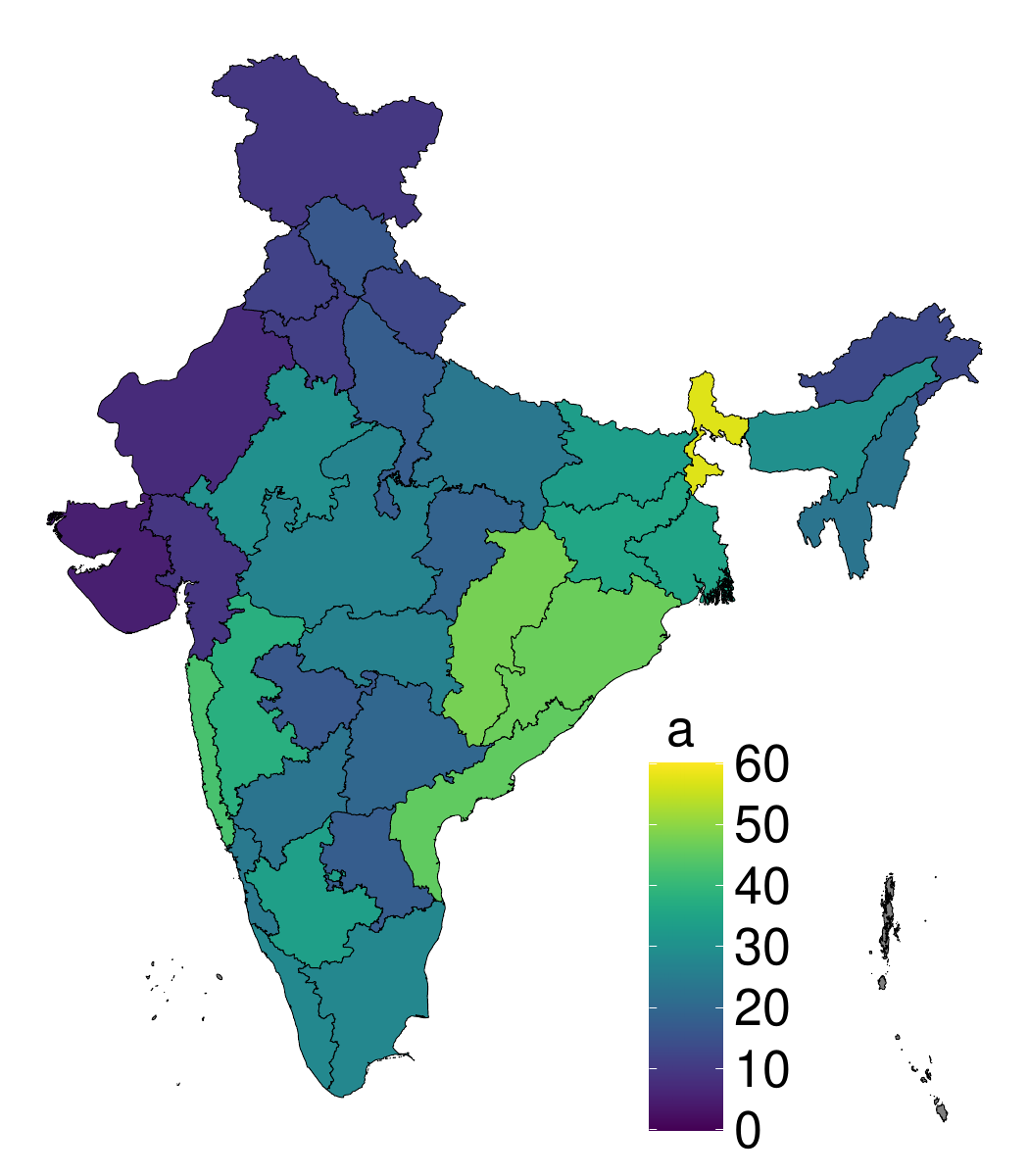}
    \includegraphics[width=0.32\textwidth]{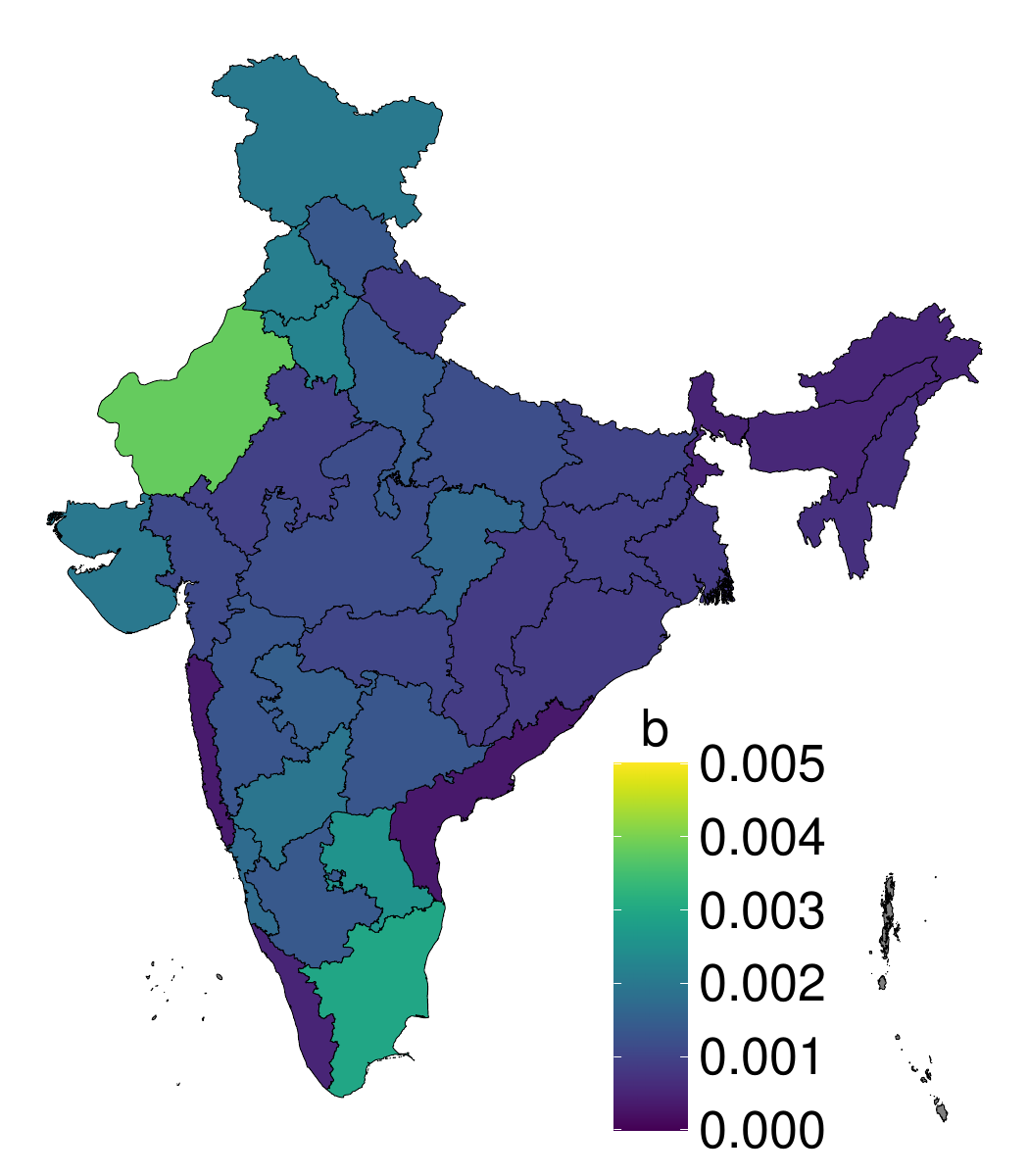}
    \includegraphics[width=0.32\textwidth]{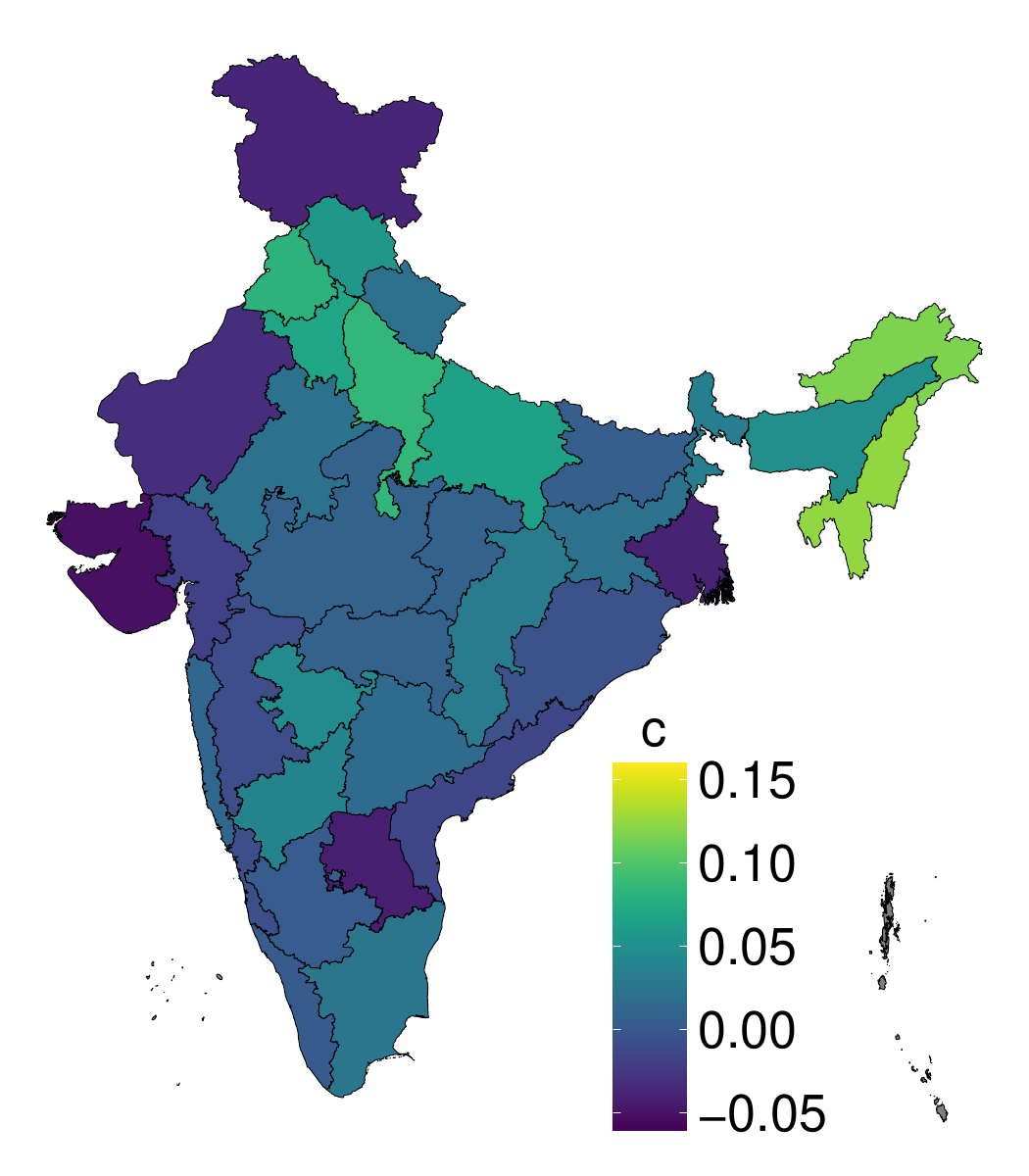}
    \caption{MLEs of the parameter vectors $\bm{a}$ (left), $\bm{b}$ (middle), and $\bm{c}$ (right) for 34 meteorological subdivisions of mainland India.}
    \label{fig:fig_mles_abc}
\end{figure}

Unlike the administrative state-wise boundaries, mainland India (excluding the Andaman and Nicobar Islands and the Lakshadweep Archipelago) comprises 34 meteorological subdivisions, which are determined based on meteorological homogeneity \citep{guhathakurta2008trends}. We present the geographical boundaries of these 34 subdivisions in Figure \ref{fig:fig_mles_abc}. We obtain monsoon (July--September) total rainfall data (in mm) for these subdivisions, covering the years 1951--2014, from the Open Government Data (OGD) Platform, India (\href{https://data.gov.in}{https://data.gov.in}), along with shape files of the boundaries. Based on the rainfall data collected from 641 districts across India by the India Meteorological Department (IMD), \cite{guhathakurta2008trends} computed monthly rainfall amounts for these districts by averaging the rainfall data from available stations within each district for each month. An area-weighted averaging method was then applied to derive subdivision-wise rainfall data from the district-wise data. \cite{guhathakurta2011impact} provide a more detailed report on the dataset preparation. The rest of the paper focuses on analyzing this dataset, which we refer to as the ``Indian rainfall dataset".

Suppose we denote the monsoon total rainfall for the $i$th meteorological subdivision and $t$th year by $Y_{it}$, for $i=1,\ldots, n$ and $t=1,\ldots, T$ with $n=34$ and $T=64$. Given the extensive $64$--year observational period encompassing phases of global warming, significant trends across the years are plausible. The data are positive-valued; hence, we can model the marginal distributions using log-normal or gamma regression, for example, where we can model the mean as a function of the year. A log-normal distribution assumption is more prevalent in the general spatial or areal data analysis literature because it allows for a logarithmic transformation of the data, enabling model fitting using an ordinary or spatial linear regression framework, rather than a generalized ordinary or spatial linear regression framework. At the same time, as mentioned in Section \ref{sec:intro}, the gamma distribution assumption for the marginal distributions is the most popular in rainfall modeling literature.

\begin{figure}[t]
    \centering
    \includegraphics[height=0.37\textwidth]{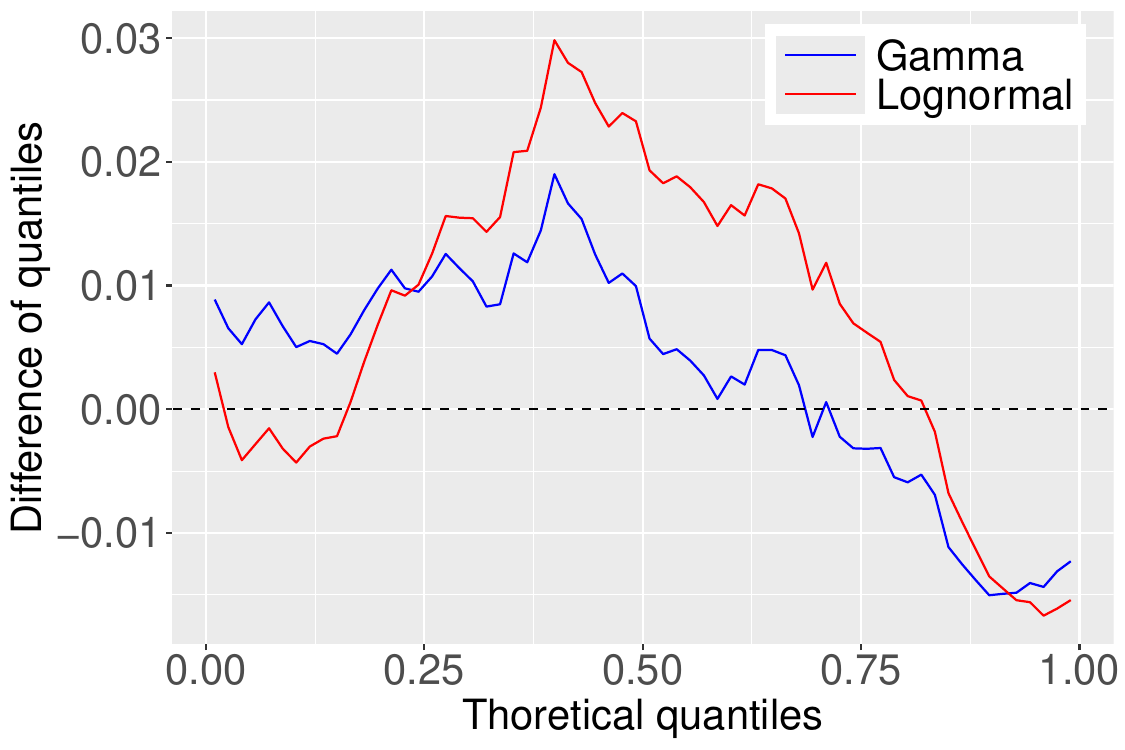}
    \includegraphics[height=0.37\linewidth]{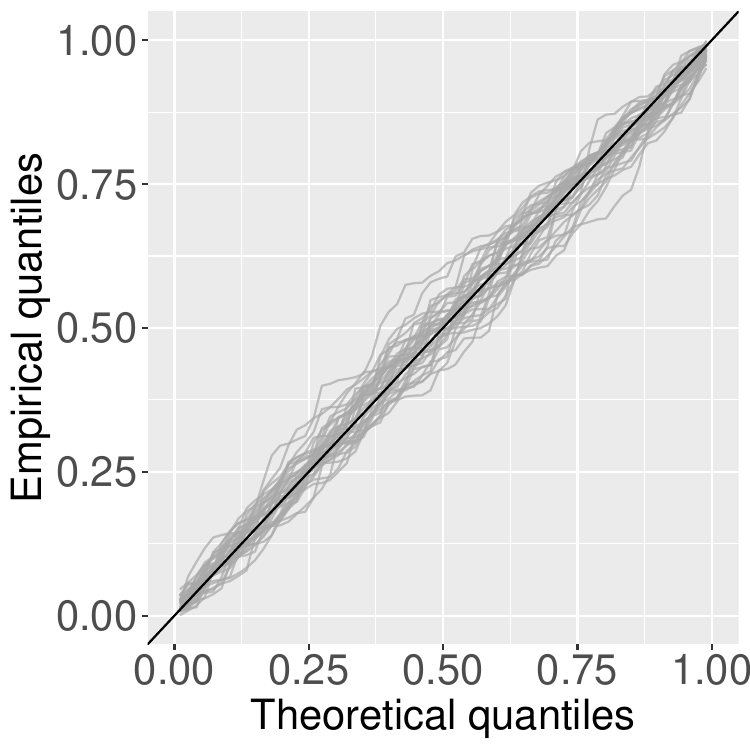}
    \caption{Left: Difference between the uniform transformed data-quantiles and $\textrm{Uniform}(0,1)$ quantiles for the fitted gamma and lognormal regression models after combining all subdivisions. Here, a model is preferred over the other if the difference in quantiles remains nearer to zero (the dashed line);  Right: Q-Q plots of the uniform-transformed data quantiles when the marginal distributions are gamma for 34 meteorological subdivisions of mainland India.}
    \label{fig:qq_plot_comparison}
\end{figure}

We first postulate that the data are independently distributed across years and marginally follow a gamma distribution. To model the space-time varying mean parameter, we write $\text{E}(Y_{it}) = \mu_{it}$ with spatially varying shape parameter $a_i$. Hence, for every region index $i$, we assume
\begin{equation} \label{eq:gamma_marginals}
    Y_{it} \overset{\mathrm{ind}}{\sim} \text{Gamma}(\mu_{it}, a_i), \ t = 1, \dots, T,
\end{equation} 
where $a_i$ is the shape parameter and $\mu_{it}$ is the mean of the gamma distribution for region index $i $ and time point $t$, which is connected with a log-link function
\begin{equation}\label{eq:gamma_glm_loglink}
    \log(\mu_{it}) = \alpha_i + \beta_i t, \ t = 1, \dots, T,
\end{equation}
in which $\alpha_i$ and $\beta_i$ are spatially varying coefficients (SVCs). A competing log-normal model would be $Y^*_{it} = \log(Y_{it}) \overset{\mathrm{ind}}{\sim} \text{Normal}(\mu_{it}, \sigma^2_i), \ t = 1, \dots, T$, where $\mu_{it} = \alpha_i^* + \beta_i^* t$ for SVCs $\alpha^*_i$, $\beta^*_i$, and variances $\sigma^2_i$. We separately fit the regression models for every region index $i$ using maximum likelihood estimates and compare their uniform-transformed QQ plots in the left panel of Figure \ref{fig:qq_plot_comparison}. The lognormal model provides a better fit for low quantile levels, while the gamma model outperforms the lognormal model overall. Compared to the lognormal model, the gamma model provides approximately 7\% lower $\textrm{RMSE}_U$ and $\textrm{MAE}_U$. For every subregion $i$, we present the uniform-transformed quantiles based on gamma regression in the right panel of Figure \ref{fig:qq_plot_comparison}. Except for certain middle-range quantiles, the model fits the data reasonably well. In the rest of the paper, we subsequently consider the gamma model in \eqref{eq:gamma_marginals} and \eqref{eq:gamma_glm_loglink}. 

We further reparameterize the gamma model in \eqref{eq:gamma_marginals} and \eqref{eq:gamma_glm_loglink} in terms of the standard shape-rate parameters $(a_i, \lambda_{it})$, as this representation clarifies the role of each parameter. The time covariate is centered and scaled to $t^*$ to reduce correlation between the intercept and slope and improve estimation stability \citep{gelman2008weakly}. 
\[t^* = \frac{t - m_t}{s_t},\quad m_t = \frac{1}{T} \sum_{t=1}^T t,\quad  s_t^2 = \frac{1}{T} \sum_{t=1}^T t^2 - m_t^2.\]
With this transformation, we may rewrite the log-link in \eqref{eq:gamma_glm_loglink} as 
\[\log(\mu_{it}) = \tilde{\alpha}_i + \tilde{\beta}_i t^*,\] 
where $\tilde{\alpha}_i = \alpha_i + \beta_i m_t$, $\tilde{\beta}_i = \beta_i s_t$.

Under the standard shape-rate gamma parameterization, we have the mean $\mu_{it} = {a_i}/{\lambda_{it}}$. Substituting the above expression of $\mu_{it}$ yields 
\[ \lambda_{it} = \frac{a_i}{\mu_{it}} = a_i \exp[-\tilde{\alpha}_i] \exp[-\tilde{\beta}_i t^*]. \]
For interpretability, let $b_i = \exp[-\tilde{\alpha}_i]$ and $c_i = -\tilde{\beta}_i$. Then the rate can be written compactly as
\begin{equation}\label{eq:rate_par_gamma}
   \lambda_{it}  = a_i b_i \exp[c_i t^*].
\end{equation}
Here, $a_i$ controls the shape (or equivalently, the coefficient of variation), $b_i$ is an intercept-related parameter, and $c_i$ is a slope-related parameter that captures the climate change trend for subdivision $i$. With this reparameterization, the mean and variance of $Y_{it}$ follow as
\begin{equation} \label{eq:interpretation}
\begin{split}
    \text{E}(Y_{it}) & = \frac{a_i}{\lambda_{it}} = b_i^{-1} \exp[-c_it^*], \\ 
    \text{Var}(Y_{it}) & = \frac{a_i}{\lambda_{it}^2} = \frac{1}{a_i}(b_i^{-1} \exp[-c_i t^*])^2.
\end{split}
\end{equation}
Thus, while $b_i$ shifts the overall rainfall level for region $i$, $c_i$ determines the temporal trend, and the coefficient of variation remains constant over time at $a_i^{-1/2}$ so that variance-to-mean ratios remain unchanged across regions. In this reparameterization, expressing the model in terms of $(a_i, \lambda_{it})$ clarifies the relationship between the mean and variance, and simplifies expressions for the variance (as in \eqref{eq:interpretation}). This decomposition provides interpretable parameters that can be meaningfully compared across different climatic regions of India.

For exploratory analysis, we fit the model in \eqref{eq:gamma_marginals} and \eqref{eq:rate_par_gamma} separately for every region $i$ using maximum likelihood estimation procedure and we present the estimates (MLEs) of $\bm{a} = (a_1, \dots, a_n)^\intercal$, $\bm{b} = (b_1, \dots b_n)^\intercal$ and $\bm{c} = (c_1, \dots, c_n)^\intercal$ in Figure \ref{fig:fig_mles_abc}. The standard errors of the estimates are low due to $T=64$ replications used for the analysis, and hence, any spatial pattern visible in Figure \ref{fig:fig_mles_abc} can be conceptualized to build a flexible model for them. First, we notice that the MLEs of $a_i$s, $b_i$s, and $c_i$s are non-constant across space, motivating us to consider SVCs. Second, the estimates for nearby regions exhibit similarity, indicating the suitability of a CAR or an ICAR prior for the parameters.

We further explore the requirement of a spatial modeling of the likelihood layer for the Indian rainfall dataset. Denoting the gamma CDF for region $i$ by $F(\cdot; a_i, b_i, c_i)$ and the corresponding MLEs of the model parameters by $\hat{a}_i$, $\hat{b}_i$, and $\hat{c}_i$, and based on the suitability of the gamma distribution for the data, $F(Y_{it}; \hat{a}_i, \hat{b}_i, \hat{c}_i)$ follows a $\textrm{Uniform}(0,1)$ distribution (approximately). Thus, $\Phi^{-1}(F(Y_{it}; \hat{a}_i, \hat{b}_i, \hat{c}_i)) = X_{it}$ (say) follows a $\textrm{Normal}(0,1)$ distribution (approximately, where $\Phi^{-1}(\cdot)$ is the standard normal quantile function) and we can assess the spatial dependence in $Y_{it}$s via the Moran's I of $X_{it}$s. We obtain the average (over $t$) Moran's I to be $0.3613$ with a negligible $p$-value, indicating a significant positive spatial association; this motivates us to choose a CAR model for the likelihood.

\section{Methodology} 
\label{sec:methodology}
In this section, we discuss our proposed Bayesian hierarchical model. We first discuss a CAR copula model and then develop the model layers using the copula.

\subsection{CAR copula}
\label{subsec:car_copula}

Let $\tilde{\bm{X}} = (\tilde{X}_1, \ldots, \tilde{X}_n)'$ be a random vector following a CAR model defined in Section \ref{subsec:car_models}, where the adjacency matrix is constructed using the neighborhood structure of the meteorological subdivisions of mainland India. Given the spatial autocorrelation parameter $\rho$ and fixed unit marginal variance ($\sigma^2=1$), the joint distribution of $\tilde{\bm{X}}$ is given by $\tilde{\bm{X}} \sim \textrm{Normal}_n(\bm{0}, [\bm{M} - \rho \bm{W}]^{-1})$, where $\bm{M}$ and $\bm{W}$ are as defined in Section \ref{subsec:car_models} and $\textrm{Normal}_n$ denotes an $n$-variate normal distribution. Suppose the diagonal elements of the matrix $[\bm{M} - \rho \bm{W}]^{-1}$ are given by $d_i$ for $i=1,\ldots, n$ and $\bm{\Delta} = \textrm{diag}(d_1, \ldots, d_n)$. Then, $\bm{\Delta}^{-1/2} \tilde{\bm{X}} \sim \textrm{Normal}_n(\bm{0}, \bm{\Delta}^{-1/2} [\bm{M} - \rho \bm{W}]^{-1} \bm{\Delta}^{-1/2})$ and its corresponding marginal distributions are $\textrm{Normal}(0,1)$. Denoting $U_i = \Phi(d_i^{-1/2} \tilde{X}_i)$ for $i=1,\ldots, n$, where $\Phi(\cdot)$ denotes the standard normal CDF, we call the joint distribution of $\bm{U} = (U_1, \ldots, U_n)'$ a CAR copula with adjacency matrix $\bm{W}$ and spatial autocorrelation parameter $\rho$. The joint CDF of $\bm{U}$ is given by 
$$C(u_1, \ldots, u_n) = \Phi_n([\Phi^{-1}(u_1), \ldots, \Phi^{-1}(u_n)]; \bm{0}, \bm{\Delta}^{-1/2} [\bm{M} - \rho \bm{W}]^{-1} \bm{\Delta}^{-1/2}),$$
where $\Phi_n(\cdot; \bm{\mu}, \bm{\Sigma})$ denotes the joint CDF of an $n$-dimensional multivariate normal distribution with mean vector $\bm{\mu}$ and covariance matrix $\bm{\Sigma}$. The corresponding copula density function is given by
$$c(u_1, \ldots, u_n) = \frac{\phi_n([\Phi^{-1}(u_1), \ldots, \Phi^{-1}(u_n)]; \bm{0}, \bm{\Delta}^{-1/2} [\bm{M} - \rho \bm{W}]^{-1} \bm{\Delta}^{-1/2})}{\prod_{i=1}^n \phi(\Phi^{-1}(u_i))},$$
where $\phi_n(\cdot; \bm{\mu}, \bm{\Sigma})$ denotes the joint density corresponding to $\Phi_n(\cdot; \bm{\mu}, \bm{\Sigma})$ and $\phi(\cdot)$ are standard normal densities.

In our implementation, the inverse of $\bm{M} - \rho \bm{W}$ is not explicitly required, except to obtain $\bm{\Delta}$ required for scaling the copula. Rather than computing the full covariance matrix, the diagonals can be computed efficiently from the Cholesky factorization of $\bm{M} - \rho \bm{W}$. Specifically, if $\bm{M} - \rho \bm{W} =\bm{LL}^\intercal$, solving $\bm{LZ}=\bm{I}$ yields $\bm{Z}=\bm{L}^{-1}$. The diagonal matrix is further obtained as $\bm{\Delta}=\textrm{diag}(\sum_i Z_{i1}^2, \ldots, \sum_i Z_{in}^2)$. This avoids the $\mathcal{O}(n^3)$ cost of a full inversion, requiring only $\mathcal{O}(n^2)$ operations.

\subsection{Data layer modeling}
\label{subsec:data_layer}

We recall the data setup from Section \ref{sec:eda}, where the monsoon total rainfall data are denoted by $Y_{it}$ for $i=1,\ldots,n=34$ meteorological subdivisions of India and for $t=1,\ldots, T=64$ years between 1951 and 2014. 

We separately model the marginal distribution and dependence structure of $Y_{it}$s in the data layer.  Following \eqref{eq:gamma_marginals} and \eqref{eq:rate_par_gamma}, given the marginal distribution-related parameters $a_i$, $b_i$, and $c_i$, the conditional distribution of $Y_{it}$ is
\begin{equation} \label{eq:gamma_marginals_hierarchical}
    Y_{it} | a_i, b_i, c_i \overset{\mathrm{ind}}{\sim} \text{Gamma}(b_i^{-1} \exp[-c_i~t^*], a_i), \ t = 1, \dots, T,
\end{equation} 
where $t^* = (t - m_t) / s_t$ with $m_t = T^{-1} \sum_{t=1}^T t$ and $s_t^2 = T^{-1} \sum_{t=1}^T t^2 - m_t^2$. Henceforth, we denote the CDF of the gamma distribution in \eqref{eq:gamma_marginals_hierarchical} as $F(\cdot; a_i, b_i, c_i)$. Thus, $U_{it} = F(Y_{it}; a_i, b_i, c_i) \sim \textrm{Uniform}(0,1)$. Let the vectors of $a_i$s, $b_i$s, and $c_i$s be denoted by $\bm{a}$, $\bm{b}$, and $\bm{c}$, respectively, as described in Section \ref{sec:eda}. Conditioning on $\bm{a}$, $\bm{b}$, and $\bm{c}$, while we assume independence across years, we model the spatial dependence of $\bm{U}_t =(U_{1t}, \ldots, U_{nt})^\intercal$ in a copula framework as described in Section \ref{subsec:car_copula}. Overall, the density of $\bm{Y}_t=(Y_{1t}, \ldots, Y_{nt})^\intercal$ given $\bm{a}$, $\bm{b}$, $\bm{c}$, and $\rho$ is 
\begin{eqnarray} \label{eq:gamma_car_joint}
   \nonumber  f_n(\bm{y}_t | \bm{a}, \bm{b}, \bm{c}, \rho) &=& \phi_n([\Phi^{-1}(u_{1t}), \ldots, \Phi^{-1}(u_{nt})]; \bm{0}, \bm{\Delta}^{-1/2} [\bm{M} - \rho \bm{W}]^{-1} \bm{\Delta}^{-1/2}) \\
    && \times \prod_{i=1}^n f(y_{it}; a_i, b_i, c_i) \big/ \prod_{i=1}^n \phi(\Phi^{-1}(u_{it}))~~,
\end{eqnarray}
where $u_{it} = F(y_{it}; a_i, b_i, c_i)$, $f(y_{it}; a_i, b_i, c_i)$ denotes the density corresponding to the gamma distribution in \eqref{eq:gamma_marginals_hierarchical}. Further, assuming independence across time, the joint density of the full data is $f_n(\bm{y}_1, \ldots, \bm{y}_T | \bm{a}, \bm{b}, \bm{c}, \rho) = \prod_{t=1}^T f_n(\bm{y}_t | \bm{a}, \bm{b}, \bm{c}, \rho)$. To assess whether the spatial dependence structure is stationary over time (i.e., whether the dependence parameter $\rho$ varies systematically across years), we compute year-specific, smoothed maximum-likelihood estimates of $\rho$ using a moving-average procedure. Smoothed maximum likelihood estimates of $\rho$ remain consistently high with no systematic temporal trend, supporting our choice of the same $\rho$ across years. The details and figure are provided in Section 2 of the Supplementary Material.

In this proposed Bayesian framework, missing rainfall observations are treated as additional unknown quantities and imputed within the MCMC sampling scheme, conditional on the observed data and model parameters. This procedure yields valid posterior inference under the assumption that the missingness mechanism is ignorable \citep{Rubin1976, little2019statistical}, which holds when data are either Missing Completely at Random (MCAR) or Missing At Random (MAR). We do not model the missingness mechanism explicitly, and thus do not accommodate Not Missing At Random (NMAR) scenarios, where the probability of missingness depends on unobserved values. In practice, MAR is the most plausible assumption for our dataset, as missing entries typically arise from administrative or instrumental recording failures rather than systematic dependence on unobserved rainfall values. We adopt the partitioning approach of \cite{rue2005gaussian} for handling missing data, which is more efficient than directly inverting the precision matrix $\bm{Q}$. The computational cost of our rainfall application with $n=34$ regions is negligible, but these techniques ensure scalability to larger spatial dimensions.

\subsection{Bayesian latent Gaussian model}
\label{subsec:bayesian_lgm}
In this subsection, we subsequently model the SVCs $a_i$s, $b_i$s, and $c_i$s. The range of these coefficients are $a_i > 0, b_i > 0, c_i \in \mathbb{R}$. We thus consider the logarithmic transformation of the first two and define $a_i^* = \log(a_i)$ and $b_i^* = \log(b_i)$ and choose different dependent or independent Gaussian priors for them as follows.
\vspace{2mm}

\noindent \textbf{Independent priors (Indep):} We assume independent and identically distributed (iid) Gaussian priors to the parameters as follows
\begin{equation}\label{eq:priors_indep}
    \begin{split}
        a_i^* | \mu_a, \sigma_a^2 &\overset{\mathrm{iid}}{\sim} \textrm{Normal}(\mu_a, \sigma_a^2),~\\
        {b}_i^* | \mu_b, \sigma_b^2 &\overset{\mathrm{iid}}{\sim} \textrm{Normal}(\mu_b, \sigma_b^2),~\\
          c_i | \mu_c, \sigma_c^2 &\overset{\mathrm{iid}}{\sim} \textrm{Normal}(\mu_c, \sigma_c^2),
    \end{split}
\end{equation}
for all $i = 1, 2, \dots, n$. Here, $\mu_a$ and $\sigma_a^2$ represent the overall average of $a_i^*$s and the spatial variability of $a_i^*$s considering mainland India. Similarly, $\mu_b$ and $\sigma_b^2$ represent the overall average of $b_i^*$s and spatial variability of $b_i^*$s, and $\mu_c$ and $\sigma_c^2$ represent the overall average of $c_i$s and spatial variability of $c_i$s. The prior choice in \eqref{eq:priors_indep} does not allow a spatial dependence structure visible from Figure \ref{fig:fig_mles_abc}. Besides, the rainfall patterns of two neighboring regions usually behave similarly for natural climatic reasons. Hence, it is likely that \eqref{eq:priors_indep} would underperform compared to a model that appropriately considers a spatial dependence structure, and the spatial smoothing of the parameters is controlled by the data only.
\vspace{2mm}

\noindent \textbf{CAR priors (CAR):} Considering the potential geographical relationships of the rainfall pattern for two nearby meteorological subdivisions, we consider CAR prior specifications for the parameters as follows
\begin{equation}\label{eq:priors_car}
    \begin{split}
        \bm{a}^* = (\log (a_1), \dots, \log (a_n) )' & {\sim} \text{Normal}_n(\mu_a \bm{1}, \sigma_a^2[\bm{M} - \rho_a\bm{W}]^{-1}),\\
        \bm{b}^* = (\log (b_1), \dots, \log (b_n) )' & {\sim} \text{Normal}_n(\mu_b \bm{1}, \sigma_b^2[\bm{M} - \rho_b\bm{W}]^{-1}),\\
          \bm{c}  = (c_1, \dots, c_n)' & {\sim} \text{Normal}_n(\mu_c \bm{1}, \sigma_c^2[\bm{M} - \rho_c\bm{W}]^{-1}),
    \end{split}
\end{equation}
where $\rho_a, \rho_b,$ and $\rho_c$ capture spatial autocorrelation of the parameters $a_i$s, $b_i$s, and $c_i$s. The other hyperparameters $\mu_a$, $\mu_b$, $\mu_c$, $\sigma_a^2$, $\sigma_b^2$, and $\sigma_c^2$ have similar meaning as in Case 1. The adjacency matrix is the same as in Section \ref{subsec:data_layer}.
\vspace{2mm}

\noindent \textbf{Intrinsic CAR priors (ICAR):} While the spatial correlation structure in \eqref{eq:priors_car} is more general, estimating the hyperparameters $\rho_a, \rho_b,$ and $\rho_c$ is usually challenging due to high posterior uncertainty. Thus, as a special case, we consider Intrinsic CAR (ICAR) \citep{freni2018note} prior specifications for the parameters as follows
\begin{equation}\label{eq:priors_icar}
    \begin{split}
        \bm{a}^* = (\log (a_1), , \dots, \log (a_n) )^\intercal & {\sim} \text{Normal}_n(\mu_a \bm{1}, \sigma_a^2[\bm{M} - \bm{W}]^{-1}),\\
        \bm{b}^* = (\log (b_1), , \dots, \log (b_n) )^\intercal & {\sim} \text{Normal}_n(\mu_b \bm{1}, \sigma_b^2[\bm{M} - \bm{W}]^{-1}),\\
          \bm{c}  = (c_1, \dots, c_n)^\intercal & {\sim} \text{Normal}_n(\mu_c \bm{1}, \sigma_c^2[\bm{M} - \bm{W}]^{-1}),
    \end{split}
\end{equation}
where we set $\rho_a=1, \rho_b=1$ and $\rho_c=1$ in Case 2. While the priors are improper due to singular covariance matrices, their posteriors are proper, and hence, the inferences are reliable.

Finally, we choose a weakly-informative $\textrm{Uniform}(0,1)$ prior for $\rho$. For the hyperparameters, we again choose weakly informative hyperpriors
\begin{equation}\label{eq:hyperpriors}
    \begin{split}
        \mu_a, \mu_b, \mu_c &\overset{\mathrm{iid}}{\sim} \textrm{Normal}(0, 10^2),~\\
        \sigma_a^2, \sigma_b^2, \sigma_c^2 &\overset{\mathrm{iid}}{\sim} \textrm{Inverse-gamma}(0.01, 0.01),~\\
          \rho_a, \rho_b, \rho_c &\overset{\mathrm{iid}}{\sim} \textrm{Uniform}(0, 1).
    \end{split}
\end{equation}

Altogether, \eqref{eq:gamma_car_joint}, \eqref{eq:priors_car}, and \eqref{eq:hyperpriors} specify a fully Bayesian hierarchical model, and we henceforth call it the CAR-CAR model due to a CAR structure at both data and prior layers. Similarly, we call the model specified by \eqref{eq:gamma_car_joint}, \eqref{eq:priors_icar}, and \eqref{eq:hyperpriors} (ignoring the hyperpriors for $\rho_a$, $\rho_b$, and $\rho_b$) as CAR-ICAR model, and the model specified by \eqref{eq:gamma_car_joint}, \eqref{eq:priors_indep}, and \eqref{eq:hyperpriors} as CAR-Indep model. For model comparison purposes, we consider several competing models by setting $\rho=0$, i.e., the independence at the data layer, and combining with different cases of prior specifications. We refer to them as Indep-Indep, Indep-CAR, and Indep-ICAR models, respectively.

\section{Bayesian computation}
\label{sec:computation}

This section discusses the MCMC sampling steps for the CAR-CAR model. Sampling for the other models mentioned in the previous section can be performed simply by ignoring specific update steps. For example, the computation for the CAR-ICAR model can be done just by ignoring the updates of $\rho_a,\rho_b, \text{and}~\rho_c$ and setting them to one.

\subsection{Gibbs sampling steps for the CAR-CAR model}

The set of parameters and hyperparameters in the model is
\begin{eqnarray}
\nonumber && \bm{\Theta} = \left\lbrace \bm{a}^*, \bm{b}^*, \bm{c}, \rho, \mu_a, \mu_b, \mu_c, \sigma_a^2, \sigma_b^2, \sigma_c^2, \rho_a, \rho_b, \rho_c, Y^*_{it}s \right\rbrace,
\end{eqnarray}
which includes parameter vectors, single parameters, and hyperparameters. We update each of the vectors $\bm{a}^*$, $\bm{b}^*$, and $\bm{c}$ simultaneously using a block MH algorithm. Here, by $Y_{it}^*$s, we denote the missing data that needs to be imputed within the MCMC steps. The update steps within a Gibbs sampler are as follows.

\begin{itemize}
    \item \textbf{Updating $\bm{a}^*$:} 
    Sample $\bm{a}^*$ via Metropolis-Hastings algorithm from 
    \begin{equation*}
    \begin{split}
        & f(\bm{a}^*|\bm{Y}_1, \ldots, \bm{Y}_T, \bm{b}^*, \bm{c}, \rho, \mu_a, \sigma_a^2, \rho_a) \\
         & \propto \prod_{t=1}^T f_n(\bm{y}_t | \bm{a}, \bm{b}, \bm{c}, \rho) \times \phi_{n}(\bm{a}^* ; \mu_a \bm{1}, \sigma_a^2[\bm{M} - \rho_a\bm{W}]^{-1}),
    \end{split}
    \end{equation*}
    where $f_n$ is given by \eqref{eq:gamma_car_joint}, $a_i=\exp[a_i^*]$, $b_i = \exp[b_i^*]$ for $i=1,\ldots, n$, $\phi_n(\cdot; \bm{\mu}, \bm{\Sigma})$ is the multivariate normal density with mean vector $\bm{\mu}$ and covariance matrix $\bm{\Sigma}$.
    \item \textbf{Updating $\bm{b}^*$:}
    Similar to $\bm{a}^*$, sample $\bm{b}^*$ jointly via Metropolis-Hastings algorithm from
    $f(\bm{b}^*|\bm{Y}_1, \ldots, \bm{Y}_T, \bm{a}^*, \bm{c}, \rho, \mu_b, \sigma_b^2, \rho_b)$.
    \item \textbf{Updating $\bm{c}$:}
    Similar to $\bm{a}^*$, and $\bm{b}^*$, sample $\bm{c}$ jointly via Metropolis-Hastings algorithm from $f(\bm{c}|\bm{Y}_1, \ldots, \bm{Y}_T, \bm{a}^*, \bm{b}^*, \rho, \mu_c, \sigma_c^2, \rho_c)$ .
    \item \textbf{Updating $\rho$:}
    Sample $\rho$ via the Metropolis-Hastings algorithm from
    \begin{equation*}
    \begin{split}
        & f(\rho|\bm{Y}_1, \ldots, \bm{Y}_T, \bm{a}^*, \bm{b}^*, \bm{c}) \\
         &  \propto \prod_{t=1}^T \phi_n([\Phi^{-1}(U_{1t}), \ldots, \Phi^{-1}(U_{nt})]; \bm{0}, \bm{\Delta}^{-1/2} [\bm{M} - \rho \bm{W}]^{-1} \bm{\Delta}^{-1/2}) \mathbbm{1}(\rho \in (0, 1)),
    \end{split}
    \end{equation*}
    where $U_{it} = F(Y_{it}; a_i=\exp[a_i^*], b_i=\exp[b_i^*], c_i)$ with $F(\cdot; a, b, c)$ the CDF of a gamma distribution as described in \eqref{eq:gamma_marginals_hierarchical}, and $\mathbbm{1}(\cdot)$ is an indicator.
    \item \textbf{Updating $\mu_a$:}
    Sample $\mu_a$ directly from the full conditional distribution, which is
    \begin{equation*}
        \mu_{a}|\bm{a}^*, \rho_a, \sigma_a^2 \sim \textrm{Normal}\left(\frac{\sigma_a^{-2} \bm{1}' [\bm{M} - \rho_a \bm{W}] \bm{a}^*}{\sigma_a^{-2} \bm{1}' [\bm{M} - \rho_a \bm{W}] \bm{1}+ 10^{-2}},\ \frac{1}{\sigma_a^{-2} \bm{1}' [\bm{M} - \rho_a \bm{W}] \bm{1}+ 10^{-2}}\right).
    \end{equation*}
    \item \textbf{Updating $\mu_b$:}
Similar to $\mu_a$, update $\mu_b$ by sampling directly from the full conditional distribution $f(\mu_{b}|\bm{b}^*, \sigma_b^2, \rho_b)$.
    \item \textbf{Updating $\mu_c$:}
Similar to $\mu_a$, update $\mu_c$ by sampling directly from the full conditional distribution $f(\mu_{c}|\bm{c}, \sigma_c^2, \rho_c)$.
    \item \textbf{Updating $\sigma_a^2$:}
    Sample $\sigma_a^2$ directly from the full conditional distribution, which is
    \begin{equation*}
        \sigma_a^2|\bm{a}^*, \mu_{a}, \rho_a \sim \text{Inverse-gamma}\left(0.5n+0.01, 0.5 \left(\bm{a}^*-\mu_a \bm{1}\right)' [\bm{M} - \rho_a \bm{W}]\left(\bm{a}^*-\mu_a \bm{1}\right) + 0.01\right)
    \end{equation*}
    \item \textbf{Updating $\sigma_b^2$:}
    Similar to $\sigma_a^2$, update $\sigma_b^2$ by sampling directly from the full conditional distribution $f(\sigma_b^2|\bm{b}^*, \mu_{b}, \rho_b)$.
    \item \textbf{Updating $\sigma_c^2$:}
    Similarly, update $\sigma_c^2$ by sampling directly from the full conditional distribution $f(\sigma_c^2|\bm{c}, \mu_{c}, \rho_c)$.
    \item \textbf{Updating $\rho_a$:}
    Sample $\rho_a$ via the Metropolis-Hastings algorithm from
    \begin{equation*}
    f(\rho_a|\bm{a}^*, \mu_a, \sigma_a^2) \propto \phi_{n}(\bm{a}^* ; \mu_a \bm{1}, \sigma_a^2[\bm{M} - \rho_a\bm{W}]^{-1}) \mathbbm{1}(\rho_a \in (0, 1)).
    \end{equation*}
    \item \textbf{Updating $\rho_b$:}
    Similarly, sample $\rho_b$ from $f(\rho_b|\bm{b}^*, \mu_b, \sigma_b^2)$ via Metropolis-Hastings algorithm as $\rho_a$.
    \item \textbf{Updating $\rho_c$:}
    Similarly, sample $\rho_b$ from $f(\rho_c|\bm{c}, \mu_c, \sigma_c^2)$ via Metropolis-Hastings algorithm as $\rho_a$ and $\rho_b$.
    \item \textbf{Updating $Y^*_{it}s$:} For a time point $t$, if {$\mathcal{I}^{(t)}$} denotes the region indexes where the data are missing, we construct $\bm{Y}_t^*$ (vector of missing values) based on $i \in \mathcal{I}^{(t)}$ and the vector for the rest of the indexes as $\widetilde{\bm{Y}}_t$, i.e., based on all $i \in \bar{\mathcal{I}}^{(t)} = \{1, 2, \dots, n\}\setminus\mathcal{I}^{(t)}$. Further, we transform all the components to $\textrm{Normal}(0,1)$ scale by the $\Phi^{-1}(F(Y_{it}; a_i, b_i, c_i))$ transformation. Given that the transformed data follow a multivariate normal distribution, one can simulate the observations corresponding to the missing locations using the standard result of the conditional distribution for multivariate normal vectors, and then transform them back to the original scale using the Gaussian CDF and the gamma quantile function. However, as obtaining the precision matrix is easier in our case, we simulate using the result of the conditional distribution for multivariate normal vectors based on the precision matrix, which allows for faster analysis \cite{rue2005gaussian}. Here, we use the partition of the precision matrix instead of inverting the parts of the variance-covariance matrix. Let the precision matrix $\bm{Q}$ of $\bm{Y}_t = ({\bm{Y}_t^*}^\intercal,\widetilde{\bm{Y}}_t^\intercal)^\intercal$ be partitioned according to the missing observations as
    \[ \bm{Q} = \begin{bmatrix}
        \bm{Q}_{{\mathcal{I}}^{(t)}, {\mathcal{I}}^{(t)}} & \bm{Q}_{{\mathcal{I}}^{(t)}, \bar{\mathcal{I}}^{(t)}} \\ \\
        \bm{Q}_{\bar{\mathcal{I}}^{(t)}, {\mathcal{I}}^{(t)}} & \bm{Q}_{\bar{\mathcal{I}}^{(t)}, \bar{\mathcal{I}}^{(t)}}
    \end{bmatrix}.
    \]
    Since the means of the marginal distributions are zero, the conditional distribution of the missing data is $\bm{Y}_t^*\mid \widetilde{\bm{Y}}_t \sim \mathrm{Normal}_{|{\mathcal{I}}^{(t)}|}\left(-\bm{Q}_{{\mathcal{I}}^{(t)},{\mathcal{I}}^{(t)}}^{-1}\bm{Q}_{{\mathcal{I}}^{(t)},\bar{\mathcal{I}}^{(t)}}\widetilde{\bm{Y}}_t,\ \bm{Q}_{{\mathcal{I}}^{(t)},{\mathcal{I}}^{(t)}}\right)$. Here $|{\mathcal{I}}^{(t)}|$ is the number of indices in the index set ${\mathcal{I}}^{(t)}$, i.e., the total number of missing observations at time point $t$.
\end{itemize}
For the univariate parameters updated via Metropolis–Hastings, alternative samplers such as the slice sampler with stepping-out \citep{neal2003slice} could also be employed, which would reduce the need for proposal tuning; we retain the Metropolis-Hastings steps here due to its superior popularity in the environmental statistics literature.

\subsection{MCMC implementation}

We implement the MCMC algorithm in \texttt{R} (\url{http://www.r-project.org}). For both simulation studies and the Indian rainfall data application, we draw 200,000 samples from the posterior distribution and discard the first 40,000 observations as burn-in. Further, we perform thinning by 20 iterations and draw inferences based on 8,000 post-burn-in samples from the posterior distribution. For the real data application, where the data includes missing observations at 3 out of 64 years, the computation time is approximately 36 minutes. In the simulation studies, which do not contain any missing data, the computation time is approximately 18 minutes. We perform all computations on a workstation equipped with an AMD Ryzen 9 5900X processor and 64GB of RAM. We monitor the convergence and mixing through Geweke statistics and effective sample sizes \citep{reich2019bayesian}.

\section{Simulation study}
\label{sec:simulation}

We conduct a simulation study to assess the accuracy of the estimation procedure outlined in Section \ref{sec:computation} and to investigate how spatial correlation can effectively reduce the uncertainty of parameter estimates. For each of the three choices of $\rho\in \{0, 0.5, 0.9 \}$, representing zero, moderate, and high spatial correlation, we simulate 100 datasets from the proposed model in Section \ref{subsec:data_layer}. Here, we set the true parameter values for $\bm{a}$, $\bm{b}$, and $\bm{c}$ to the estimates obtained from the real dataset in Figure \ref{fig:fig_mles_abc}. Here, we compare models having a CAR structure at the data level and either independence, CAR, or ICAR structure at the prior level. For all values of $\rho$ in the data-generating models, the marginal distribution of the observations remains the same for each region due to choosing the same spatially varying coefficient (SVC) vectors. We adopt certain evaluation metrics discussed in Section \ref{subsec:eval_metrics} since they collectively capture complementary aspects of the model performance: estimation accuracy, posterior uncertainty, and validity of frequentist properties.


\subsection{Evaluation criteria}
\label{subsec:eval_metrics}
To evaluate and compare models, we use three complementary metrics: mean squared error (MSE), posterior standard deviation (SD), and coverage probability (CovP). Let $\theta$ denote a generic scalar parameter of interest with true value $\theta^*$. Suppose we run $S$ independent simulation replicates. For replicate $s\in \{1,\dots,S\}$ we obtain $M$ posterior draws
$\{\theta^{(s)}_1, \dots, \theta^{(s)}_M\}$ (after burn-in and thinning). Suppose we denote the posterior mean and posterior standard deviation in replicate $s$ by
\[
\hat{\theta}^{(s)} \;=\; \frac{1}{M}\sum_{m=1}^M \theta^{(s)}_m,
\qquad
\mathrm{SD}^{(s)}(\theta) \;=\; \sqrt{\frac{1}{M-1}\sum_{m=1}^M\big(\theta^{(s)}_m - \hat{\theta}^{(s)}\big)^2 }.
\]
\begin{itemize}
    \item \textbf{Mean Squared Error (MSE):} For a parameter $\theta$ with its estimate $\hat{\theta}$, the MSE is $E[(\hat{\theta}-\theta)^2]$. In practice, for the simulation study, for each parameter $\theta$, with true value $\theta^*$ and posterior mean estimates $\{\hat{\theta}^{(1)}, \dots, \hat{\theta}^{(S)}\}$, we compute
    \[\mathrm{MSE}(\theta) = \frac{1}{S}\sum_{s=1}^S (\hat{\theta}^{(s)} - \theta^*)^2.\]
    This metric captures the overall estimation accuracy by combining bias and variance, and is standard in simulation-based model comparisons. Smaller MSE values are preferred, and comparisons of MSE between models indicate which estimator tends to be closer (in squared-error) to the true value on average.
    
    \item \textbf{Posterior Standard Deviation (SD):} In a Bayesian context, the posterior SD reflects the uncertainty in parameter estimates. Comparing average posterior SDs across models provides insight into which model yields more stable and precise inference. For each replicate, we compute the posterior SD, $\mathrm{SD}^{(s)}(\theta)$ from the $M$ posterior draws as above, and then summarize uncertainty across replicates by the average
    \[
    \overline{\mathrm{SD}}(\theta) \;=\; \frac{1}{S}\sum_{s=1}^S \mathrm{SD}^{(s)}(\theta).
    \]
    A smaller $\overline{\mathrm{SD}}$ indicates that the model typically yields more precise (less uncertain) posterior estimates. A small posterior SD is desirable only when intervals remain well calibrated; a very small SD combined with poor coverage indicates overconfident inference.
    
    \item \textbf{Coverage Probability (CovP):} The frequentist coverage of Bayesian credible intervals is assessed by checking the proportion of how many times the true value $\theta^*$ of the parameter falls within the posterior $100(1-\alpha)\%$ credible interval across repeated simulations. The coverage probability is the proportion of replicates (out of $S$) that satisfy this condition. For each replicate $s$ form a two-sided $100(1-\alpha)\%$ posterior credible interval for $\theta$, e.g. using the $\alpha/2$ and $1-\alpha/2$ posterior quantiles, $\big(L^{(s)}(\theta),\, U^{(s)}(\theta)\big) 
    \;=\; \big(q_{\alpha/2}(\{\theta^{(s)}_m\}),\; q_{1-\alpha/2}(\{\theta^{(s)}_m\})\big)$. We then define the indicator $C^{(s)}(\theta) =  1$ if $\theta^*\in \big(L^{(s)}(\theta),U^{(s)}(\theta)\big)$, and zero otherwise. Further, the coverage probability for parameter $\theta$ is
  \[
    \mathrm{CovP}(\theta) \;=\; \frac{1}{S}\sum_{s=1}^S C^{(s)}(\theta).
  \]
  For a nominal $95\%$ interval with $\alpha = 0.05$, we expect $\mathrm{CovP}$ close to $0.95$. A larger CovP is preferred because it indicates that posterior credible intervals are well-calibrated in a frequentist sense. Usually, a $\mathrm{CovP}\ll 0.95$ indicates undercoverage (overconfident) and $\mathrm{CovP}\gg 0.95$ indicates overcoverage (possibly inefficient).
\end{itemize}

In summary, MSE measures estimation accuracy by penalizing squared deviations between parameter estimates and their actual values. Posterior SD summarizes the uncertainty in Bayesian parameter estimates. CovP assesses the calibration of Bayesian credible intervals in a frequentist sense. These criteria provide a balanced framework for evaluating estimation accuracy and uncertainty quantification, central to hierarchical Bayesian modeling.

\subsection{Results}
\label{subsec:simulation_results}

For each model, we compare the MSE, posterior SD, and CovP of the parameters $\bm{a},\bm{b},\bm{c}$ and $\rho$ based on the posterior samples drawn using MCMC. For each of the 100 datasets, we obtain MSEs and posterior SDs for each $a_i$, $b_i$, and $c_i$. Here, we report the average of the MSEs for all 34 $a_i$s and 100 datasets, for example. For $\rho$, we report the average of the MSEs for all 100 datasets. We also obtain the SDs reported here via a similar averaging. For calculating CovPs, we first calculate the 95\% credible interval for the corresponding parameters and check whether the true value falls within that interval. Thus, we obtain a binary value for each dataset and each $i$. Further, we average across all 34 regions and 100 datasets for $a_i$s, $b_i$s, and $c_i$s and report the average. For $\rho$, we only obtain a similar binary value from each dataset and report the average across 100 datasets. We report the results in Table \ref{table:sim_results}. When the true data-generating model has $\rho=0$, we leave the CovPs blank for $\rho$ because the posterior distribution is supported over $(0,1)$ and does not include zero.

When true $\rho=0$, the Independent data layer models yield relatively larger MSEs and smaller coverage probabilities across all parameters, whereas the CAR data layer models clearly improve estimation accuracy. In particular, comparing all the metrics across models for $a_i$, $b_i$, and $c_i$s, we observe that the CAR-ICAR model provides the smallest MSE and the smallest SD with reasonable CovPs. Since $\rho$ is not present in the independent data layer models, we do not report those values for $\rho$. We observe similar MSE and SD values for $\rho$ with different priors with CAR data layer models. In this case, the CAR model with an ICAR prior is more suitable due to the natural spatial pattern of the true SVCs obtained from the Indian rainfall dataset. Here, the posterior variances of the parameters $\rho_a$, $\rho_b$, and $\rho_c$ are high. Hence, even if an ICAR prior is less flexible (due to fixing the spatial autocorrelation parameter to one), it provides a better model fitting. When true $\rho=0.5$, i.e., when the spatial correlation at the data layer is moderate, we see a similar performance comparison. CAR data layer models improve substantially upon Independent models. For $a_i$ and $b_i$, the CAR-ICAR model again provides the most favorable bias–variance trade-off. For $c_i$, the Independent data layer models perform slightly better, showing slightly smaller MSE and posterior SD, although comparable with the CAR-ICAR model. For $\rho$, in terms of MSE, CAR-ICAR models perform well, while for the CAR-Indep model, the average posterior SD is the smallest. When true $\rho=0.9$, i.e., when the spatial correlation at the data layer is strong, Independent models fail to recover the strong spatial correlation. In contrast, the CAR–ICAR and CAR–CAR models consistently outperform them in terms of MSE and SD across all regression parameters, except for the low coverage probability (0.9) for $\rho$ in the case of CAR-ICAR and CAR-CAR models. Based on a global comparison with all data-generating scenarios, the CAR-ICAR model is parsimonious and performs equally or better than the alternatives.

Table \ref{table:sim_DIC_WAIC} reports the average DIC and WAIC values, along with their standard errors, across all simulation replicates. As expected, models with a CAR specification at the data layer consistently achieve lower DIC and WAIC than their Independent counterparts, reflecting the benefit of explicitly modeling spatial correlation among observations. Within the Independent data layer, the choice of prior (Indep, ICAR, or CAR) produces only minor differences, and the CAR data layer models uniformly outperform all such models. Among the latter, the CAR-ICAR model attains the smallest DIC and WAIC in most cases, indicating that combining a CAR copula at the data layer with an ICAR prior on the spatially varying coefficients offers the best trade-off between fit and complexity. The CAR-CAR model exhibits comparable performance, whereas the CAR-Indep model performs slightly worse, particularly in scenarios with stronger spatial dependence. Taken together, the DIC and WAIC results reinforce the earlier findings based on MSE and coverage: the CAR-ICAR model is the most parsimonious and best-performing specification overall.

These findings highlight that the novelty of our approach lies not only in the new combination of gamma regression with CAR copula dependence, but also in its ability to achieve lower estimation uncertainty and better coverage than alternative models that ignore spatial dependence. Thus, the proposed framework offers tangible improvements in statistical performance, which makes it attractive to practitioners beyond the rainfall context.

\begin{table}[]
\caption{Average MSE, posterior SD, and coverage probability (CovP) of the SVCs and the spatial autocorrelation parameter for independent, CAR, and ICAR priors along with Indep model and CAR model specifications for the data layer. Here, we consider low through high values of the true spatial autocorrelation parameter and report the results averaged across different regions and 100 simulated datasets (for SVCs) and for $\rho$ by averaging across 100 simulated datasets. A smaller MSE and SD value and a higher CovP value are preferred.}
\label{table:sim_results}
\begin{tabular}{llllllllllll} \hline
                                                        \multicolumn{3}{c} {\begin{tabular}{c}
                                                          Latent \\
                                                          layer
                                                       \end{tabular} $\longrightarrow$} & \multicolumn{3}{c}{Indep}        & \multicolumn{3}{c}{ICAR}         & \multicolumn{3}{c}{CAR}          \\ \hline
                                                      $\rho^{\text{true}}$& {\hspace{-3mm}\begin{tabular}{l}
                                                          Data \\
                                                          layer
                                                       \end{tabular} $\downarrow$} & & MSE*& SD**& CovP & MSE*& SD**& CovP                  & MSE*& SD**& CovP                  \\ \hline
 \multicolumn{1}{|r|}{\multirow{3}{*}{0}} &   \multicolumn{1}{l|}{\multirow{3}{*}{{Indep}}}& \multicolumn{1}{l|}{$\bm{a}$} &     48.35&    4.41& \multicolumn{1}{l|}{0.96} &     43.48&    4.17& \multicolumn{1}{l|}{0.96} &     44.71&    4.25& \multicolumn{1}{l|}{0.97} \\
 
 \multicolumn{1}{|l|}{} &                          \multicolumn{1}{l|}{}    & \multicolumn{1}{l|}{$\bm{b}$} &     1.66&    2.79& \multicolumn{1}{l|}{0.95} &     1.66&    2.78& \multicolumn{1}{l|}{0.95} &     1.66&    2.79& \multicolumn{1}{l|}{0.95} \\
 
 \multicolumn{1}{|l|}{} &                          \multicolumn{1}{l|}{}    & \multicolumn{1}{l|}{$\bm{c}$} &     1.16&    2.31& \multicolumn{1}{l|}{0.92} &     1.07&    2.20& \multicolumn{1}{l|}{0.98} &     1.11&    2.27& \multicolumn{1}{l|}{0.97} \\ \hline
 
 \multicolumn{1}{|r|}{\multirow{3}{*}{0.5}} & \multicolumn{1}{l|}{\multirow{3}{*}{{Indep}}}& \multicolumn{1}{l|}{$\bm{a}$} &     47.47&    4.40& \multicolumn{1}{l|}{0.96} &     42.40&    4.15& \multicolumn{1}{l|}{0.94} &     43.73&    4.24& \multicolumn{1}{l|}{0.94} \\
 
 \multicolumn{1}{|l|}{} &                           \multicolumn{1}{l|}{}   & \multicolumn{1}{l|}{$\bm{b}$} &     1.71&    2.79& \multicolumn{1}{l|}{0.95} &     1.71&    2.79& \multicolumn{1}{l|}{0.94} &     1.71&    2.79& \multicolumn{1}{l|}{0.94} \\
 
 \multicolumn{1}{|l|}{} &                           \multicolumn{1}{l|}{}   & \multicolumn{1}{l|}{$\bm{c}$} &     1.18&    2.29& \multicolumn{1}{l|}{0.86} &     1.11&    2.17& \multicolumn{1}{l|}{0.91} &     1.14&    2.25& \multicolumn{1}{l|}{0.91} \\ \hline
 
 \multicolumn{1}{|r|}{\multirow{3}{*}{0.9}} & \multicolumn{1}{l|}{\multirow{3}{*}{{Indep}}}& \multicolumn{1}{l|}{$\bm{a}$} &     47.88&    4.40& \multicolumn{1}{l|}{0.96} &     43.65&    4.15& \multicolumn{1}{l|}{0.95} &     44.69&    4.23& \multicolumn{1}{l|}{0.96} \\
 
 \multicolumn{1}{|l|}{} &                        \multicolumn{1}{l|}{}      & \multicolumn{1}{l|}{$\bm{b}$} &     1.78&    2.79& \multicolumn{1}{l|}{0.89} &     1.78&    2.79& \multicolumn{1}{l|}{0.89} &     1.78&    2.79& \multicolumn{1}{l|}{0.90} \\
 
 \multicolumn{1}{|l|}{} &              \multicolumn{1}{l|}{}                & \multicolumn{1}{l|}{$\bm{c}$} &     1.19&    2.29& \multicolumn{1}{l|}{0.91} &     1.20&    2.13& \multicolumn{1}{l|}{0.93} &     1.19&    2.22& \multicolumn{1}{l|}{0.92} \\ \hline \hline
 
 \multicolumn{1}{|r|}{\multirow{4}{*}{0}} &   \multicolumn{1}{l|}{\multirow{4}{*}{{CAR}}}& \multicolumn{1}{l|}{$\bm{a}$} &     48.28&    4.41& \multicolumn{1}{l|}{0.95} &     43.45&    4.16& \multicolumn{1}{l|}{0.95} &     44.70&    4.24& \multicolumn{1}{l|}{0.95} \\
 
 \multicolumn{1}{|l|}{} &              \multicolumn{1}{l|}{}                & \multicolumn{1}{l|}{$\bm{b}$} &     1.67&    2.79& \multicolumn{1}{l|}{0.95} &     1.65&    2.78& \multicolumn{1}{l|}{0.95} &     1.66&    2.79& \multicolumn{1}{l|}{0.95} \\
 
 \multicolumn{1}{|l|}{} &                 \multicolumn{1}{l|}{}             & \multicolumn{1}{l|}{$\bm{c}$} &     1.16&    2.31& \multicolumn{1}{l|}{0.95} &     1.07&    2.20& \multicolumn{1}{l|}{0.95} &     1.12&    2.28& \multicolumn{1}{l|}{0.95} \\
 
 \multicolumn{1}{|l|}{} &              \multicolumn{1}{l|}{}                & \multicolumn{1}{l|}{$\rho$}   &     5.33&    4.06& \multicolumn{1}{l|}{-} &     5.11&    4.01& \multicolumn{1}{l|}{-} &     5.19&    4.01& \multicolumn{1}{l|}{-} \\ \hline
 
 \multicolumn{1}{|r|}{\multirow{4}{*}{0.5}} & \multicolumn{1}{l|}{\multirow{4}{*}{{CAR}}}& \multicolumn{1}{l|}{$\bm{a}$} &     46.17&    4.32& \multicolumn{1}{l|}{0.94} &     41.64&    4.11& \multicolumn{1}{l|}{0.95} &     42.74&    4.18& \multicolumn{1}{l|}{0.95} \\
 
 \multicolumn{1}{|l|}{} &              \multicolumn{1}{l|}{}                & \multicolumn{1}{l|}{$\bm{b}$} &     1.72&    2.79& \multicolumn{1}{l|}{0.95} &     1.71&    2.79& \multicolumn{1}{l|}{0.95} &     1.71&    2.79& \multicolumn{1}{l|}{0.95} \\
 
 \multicolumn{1}{|l|}{} &               \multicolumn{1}{l|}{}               & \multicolumn{1}{l|}{$\bm{c}$} &     1.22&    2.27& \multicolumn{1}{l|}{0.93} &     1.15&    2.27& \multicolumn{1}{l|}{0.94} &     1.18&    2.30& \multicolumn{1}{l|}{0.95} \\
 
 \multicolumn{1}{|l|}{} &              \multicolumn{1}{l|}{}                & \multicolumn{1}{l|}{$\rho$}   &     4.11&    4.47& \multicolumn{1}{l|}{0.94} &     4.11&    4.49& \multicolumn{1}{l|}{0.96} &     4.11&    4.49& \multicolumn{1}{l|}{0.96} \\ \hline
 
 \multicolumn{1}{|r|}{\multirow{4}{*}{0.9}} & \multicolumn{1}{l|}{\multirow{4}{*}{{CAR}}}& \multicolumn{1}{l|}{$\bm{a}$} &     40.09&    3.99& \multicolumn{1}{l|}{0.94} &     37.30&    3.87& \multicolumn{1}{l|}{0.95} &     37.97&    3.90& \multicolumn{1}{l|}{0.95} \\
 
 \multicolumn{1}{|l|}{} &          \multicolumn{1}{l|}{}                    & \multicolumn{1}{l|}{$\bm{b}$} &     1.80&    2.78& \multicolumn{1}{l|}{0.94} &     1.77&    2.78& \multicolumn{1}{l|}{0.94} &     1.78&    2.78& \multicolumn{1}{l|}{0.94} \\
 
 \multicolumn{1}{|l|}{} &          \multicolumn{1}{l|}{}                    & \multicolumn{1}{l|}{$\bm{c}$} &     1.19&    2.24& \multicolumn{1}{l|}{0.93} &     1.20&    2.40& \multicolumn{1}{l|}{0.96} &     1.17&    2.34& \multicolumn{1}{l|}{0.95} \\
 
 \multicolumn{1}{|l|}{} &        \multicolumn{1}{l|}{}                      & \multicolumn{1}{l|}{$\rho$}   &     0.55&    1.80& \multicolumn{1}{l|}{0.94} &     0.55&    1.80& \multicolumn{1}{l|}{0.90} &     0.55&    1.80& \multicolumn{1}{l|}{0.90} \\ \hline
\end{tabular}
\begin{tablenotes}
       \item [*] multiplied by $1000$ for MSE of the parameters $\bm{b}, \bm{c} \text{ and } \rho$.
       \item [**] multiplied by $100$ for SD of the parameters $\bm{b}, \bm{c} \text{ and } \rho$.
     \end{tablenotes}
\end{table}

\begin{table}[]
\caption{Average DIC and WAIC and their standard error (SE, reported within brackets) for independent, CAR, and ICAR priors, along with Independent and CAR model specification for the data layer. Here, we consider low through high values of the true spatial autocorrelation parameter and report the results averaged across 100 simulated datasets. A smaller DIC and WAIC with a smaller SE value are preferred.}
\label{table:sim_DIC_WAIC}
\begin{tabular}{llllll} \hline
 & Latent layer& $\longrightarrow$& Indep           & ICAR            &CAR             \\
\hline
                       $\rho^{\text{true}}$& Data layer $\downarrow$&   Criterion        & & & \\ \hline
\multirow{2}{*}{0}   & \multirow{2}{*}{Indep} & DIC& 28931.2 (6.73)  & 28925.9 (6.72)& 28926.8 (6.73)\\
                        &                      & WAIC& 28931.9 (6.73)& 28926.6 (6.73)& 28927.4 (6.73)\\ \hline

\multirow{2}{*}{0}   & \multirow{2}{*}{CAR} & DIC& 28932.3 (6.73)& 28927.0 (6.72)& 28928.1 (6.72)\\
                        &                      & WAIC& 28933.1 (6.73)& 28927.6 (6.72)  & 28928.7 (6.73)\\ \hline
                        
\multirow{2}{*}{0.5} & \multirow{2}{*}{Indep} & DIC& 28936.2 (6.94)& 28930.5 (6.93)& 28931.5 (6.93)\\
                        &                      & WAIC& 28936.9 (6.96)  & 28932.5 (6.95)& 28932.9 (6.94)\\ \hline 
                        
\multirow{2}{*}{0.5} & \multirow{2}{*}{CAR} & DIC& 28844.0 (6.67)& 28839.4 (6.66)& 28840.1 (6.67)\\
                        &                      & WAIC& 28844.5 (6.69)& 28839.9 (6.68)& 28840.7 (6.68)\\ \hline
                        
\multirow{2}{*}{0.9} & \multirow{2}{*}{Indep} & DIC& 28932.6 (9.70)& 28924.9 (9.66)& 28926.7 (9.67)  \\
                        &                      & WAIC& 28935.1 (9.74)  & 28931.4 (9.73)& 28931.6 (9.73)\\ \hline
                        
\multirow{2}{*}{0.9} & \multirow{2}{*}{CAR} & DIC& 28222.1 (6.87)  & 28217.6 (6.84)& 28218.1 (6.86)\\
                        &                      & WAIC& 28222.9 (6.86)& 28218.5 (6.83)  & 28218.8 (6.85)\\ \hline
\end{tabular}
\end{table}

\section{Indian rainfall data analysis}
\label{sec:application}
We first fit all the models (Indep-Indep, Indep-CAR, Indep-ICAR, CAR-Indep, CAR-CAR, and CAR-ICAR) to the Indian rainfall dataset and compare the models based on specific metrics. Here, unlike the simulation settings, the true values of the parameters are not known, and hence, we cannot compare the models in terms of MSEs and coverage probabilities. Hence, we only compare the average posterior SDs for $a_i$s, $b_i$s, and $c_i$s and the posterior SD of $\rho$. Besides, we compare the DIC and WAIC of the models as described in Section \ref{subsec:dic_waic}. We report the results in Table \ref{table:realdata_results}. Except for the average posterior SD of $c_i$s, all other metrics indicate similar or superior performance of the CAR-ICAR model. Regarding DIC and WAIC, the ICAR prior provides a better fitting and prediction performance than the CAR prior, followed by the independent priors; this result indicates the suitability of allowing SVCs in a Gaussian graphical model framework. The observed differences in DIC (exceeding 5) \citep{spiegelhalter2002bayesian} and WAIC (exceeding 11) are regarded as practically meaningful \citep{Gelman2014understanding}, thereby further supporting the preference for the CAR-ICAR specification. In contrast, the CAR-CAR model provides a similar fit but with higher uncertainty for a few parameters, suggesting possible overparameterization.

\begin{table}[h]
\caption{Average Standard deviations of the posterior of the parameters $\bm{a}, \bm{b}, \bm{c}$ and $\rho$ (if applicable), along with model comparison. A smaller value of the metrics indicates a better fitting and prediction performance in a leave-one-out cross-validation.}
\label{table:realdata_results}
\begin{tabular}{lllllll}\toprule
                                                                               Model& \multicolumn{1}{c}{$\frac{1}{n}\sum_i \mathrm{sd}(a_i)$} & \multicolumn{1}{c}{$\frac{1}{n}\sum_i \mathrm{sd}(b_i)$} & \multicolumn{1}{c}{$\frac{1}{n}\sum_i \mathrm{sd}(c_i)$} & \multicolumn{1}{c}{$\mathrm{sd}(\rho)$} & \multicolumn{1}{c}{DIC} & \multicolumn{1}{c}{WAIC}\\\midrule
 {Indep-Indep}& {4.447}& {2.722}& {1.997}& {-}& {29042}&{29049}\\
 {Indep-ICAR}& {4.207}& {2.719}& {1.865}& {-}& {29033}&{29044}\\
 {Indep-CAR}& {4.301}& {2.701}& {1.939}& {-}& {29037}&{29045}\\

 \multicolumn{1}{l}{{CAR-Indep}}  & \multicolumn{1}{l}{4.105}                      & \multicolumn{1}{l}{2.753}                      & \multicolumn{1}{l}{2.077}                      & \multicolumn{1}{l}{1.225}      & \multicolumn{1}{l}{{28141}}            & \multicolumn{1}{l}{{28155}} \\
 \multicolumn{1}{l}{{CAR-ICAR}} & \multicolumn{1}{l}{3.887}                      & \multicolumn{1}{l}{2.786}                      & \multicolumn{1}{l}{2.293}                      & \multicolumn{1}{l}{1.141}      & \multicolumn{1}{l}{{28133}}         & \multicolumn{1}{l}{{28144}} \\
 \multicolumn{1}{l}{{CAR-CAR}}  & \multicolumn{1}{l}{3.983}                      & \multicolumn{1}{l}{2.844}                      & \multicolumn{1}{l}{2.210}                       & \multicolumn{1}{l}{1.198}     & \multicolumn{1}{l}{{28136}}             & \multicolumn{1}{l}{{28148}}\\ \bottomrule
\end{tabular}
\begin{tablenotes}
       \item [*] multiplied by $100$ for the parameters $\bm{b}, \bm{c} \text{ and } \rho$.
\end{tablenotes}
\end{table}

To assess absolute model fit in addition to relative model selection criteria, we perform posterior predictive checks \citep{gelman1996posterior}. For each posterior draw of the parameters, we simulate replicated datasets from the posterior predictive distribution and compute a set of discrepancy statistics: overall mean, standard deviation (SD), minimum, and maximum, average of site-specific means, standard deviations, minimums and maximums, and average neighbor correlation across years. Posterior predictive $p$-values \citep{meng1994posterior} are calculated as the proportion of replicated statistics more extreme than the observed. These posterior predictive $p$-values are reported in Table~\ref{table:posterior_predictive_p_val}. Across all models, the $p$-values for the mean, SDs, minimum, maximum, site-level summaries, and neighbor correlation range between 0.14 and 0.95, indicating no evidence of lack of fit for these global summaries. The CAR-ICAR model remains the best-supported specification, combining superior relative fit (WAIC/DIC) with adequate posterior predictive performance for key distributional and spatial summaries.

\begin{table}[]
\caption{Posterior predictive $p$-value analysis for different models using different types of test statistics. A $p$-value more than $0.95$ or less than $0.05$ indicates that the model misfits the data. In an ideal case, the $p$-value is approximately $0.5$. In the bottom, we reported the coverage probability of the data, which measures the proportion of how many times the observed data falls within the $95\%$ posterior predictive data samples.}
\label{table:posterior_predictive_p_val}
\begin{tabular}{|l|cccccc|}
\hline
Test statistic $\downarrow$                                        & \begin{tabular}[c]{@{}c@{}}Indep-\\ Indep\end{tabular} & \begin{tabular}[c]{@{}c@{}}Indep-\\ ICAR\end{tabular} & \begin{tabular}[c]{@{}c@{}}Indep-\\ CAR\end{tabular} & \begin{tabular}[c]{@{}c@{}}CAR-\\ Indep\end{tabular} & \begin{tabular}[c]{@{}c@{}}CAR-\\ ICAR\end{tabular} & \begin{tabular}[c]{@{}c@{}}CAR-\\ CAR\end{tabular} \\ \hline
Mean                                                               & 0.58& 0.56& 0.58& 0.52& 0.66& 0.61\\
SD                                                                 & 0.55& 0.55& 0.55& 0.80& 0.89& 0.86\\
Minimum                                                            & 0.67& 0.65& 0.65& 0.85& 0.85& 0.85\\
Maximum                                                            & 0.36& 0.37& 0.37& 0.53& 0.61& 0.59\\ 
Site wise mean                                                     & 0.53& 0.51& 0.53& 0.50& 0.64& 0.59\\ 
Site wise SD                                                       & 0.67& 0.68& 0.68& 0.91& 0.95& 0.93\\ 
Site wise minimum       & 0.29& 0.28& 0.29& 0.14& 0.17& 0.17\\
Site wise maximum       & 0.43& 0.43& 0.43& 0.71& 0.83& 0.78\\ 
Avg neighbor correlation & -& -& -& 0.64& 0.73& 0.66\\ \hline
Coverage Probability     & 0.96& 0.96                                                  & 0.96                                                 & 0.95                                                 & 0.95                                                & 0.96                                               \\ \hline
\end{tabular}
\end{table}

We further consider the CAR-ICAR model only for drawing inferences and report the results based on it. To assess the quality of MCMC chains, we utilize Geweke statistics for the convergence of the post-burn-in samples and the effective sample size (ESS) for mixing and sample autocorrelation. We report the average of the Geweke statistics, and we use the batch-mean methodology \cite{brooks2011handbook} to obtain the ESS values for SVCs. A high ESS value is preferred. The Geweke statistic is a $z$-score; hence, its absolute value being lower than 1.96 indicates the convergence of the MCMC chains. We report the results in Table \ref{table:realdata_mcmc}. While the ESS values are relatively higher for all the model hyperparameters, they are smaller for the model parameter $\rho$. Overall, the ESS values are reasonable for all parameters and hyperparameters. Further, the Geweke statistics for all parameters and hyperparameters are close to zero and not outside the interval $\pm 1.96$, indicating convergence of the MCMC chains after the burn-in periods. Altogether, both metrics suggest that the MCMC samples are of reasonably good quality for drawing statistical inferences. Convergence of the MCMC algorithm is assessed using univariate trace plots of representative parameters across all models. The chains display stable mixing and no discernible trends. The complete set of trace plots and the ESS table of the parameters, considering all other models, are reported in Section 3 of the Supplementary Material. For the Indian monsoon data, posterior predictive checks similarly suggest that the CAR-Copula models provide a satisfactory fit. The $p$-values remain in reasonable ranges and do not indicate a systematic lack of fit. Section 4 of Supplementary Material includes further posterior predictive plots and diagnostics supporting these findings.

\begin{table}[h]
\caption{Average ESS calculated using batch-means and Geweke statistics of different parameters of the CAR-ICAR model}
\label{table:realdata_mcmc}
\begin{tabular}{lllllllllll}
                                            & $\bm{a}$ & $\bm{b}$ & $\bm{c}$ & $\rho$ & $\mu_a$ & $\mu_b$ & $\mu_c$ & $\sigma_a^2$ & $\sigma_b^2$ & $\sigma_c^2$                \\ \hline
\multicolumn{1}{|l|}{ESS} & 819& 900& 1750& 241& 8000& 8000& 8000& 1952& 8000& \multicolumn{1}{l|}{992} \\
\multicolumn{1}{|l|}{Geweke}     & -0.52& 0.65& 0.73& 0.32& -1& 0.16& -0.19& -0.75& 1.11& \multicolumn{1}{l|}{-0.07}  \\ \hline
\end{tabular}
\end{table}

\begin{figure}
    \centering
    \begin{subfigure}[b]{0.3\textwidth}
        \centering
        \includegraphics[width=\textwidth]{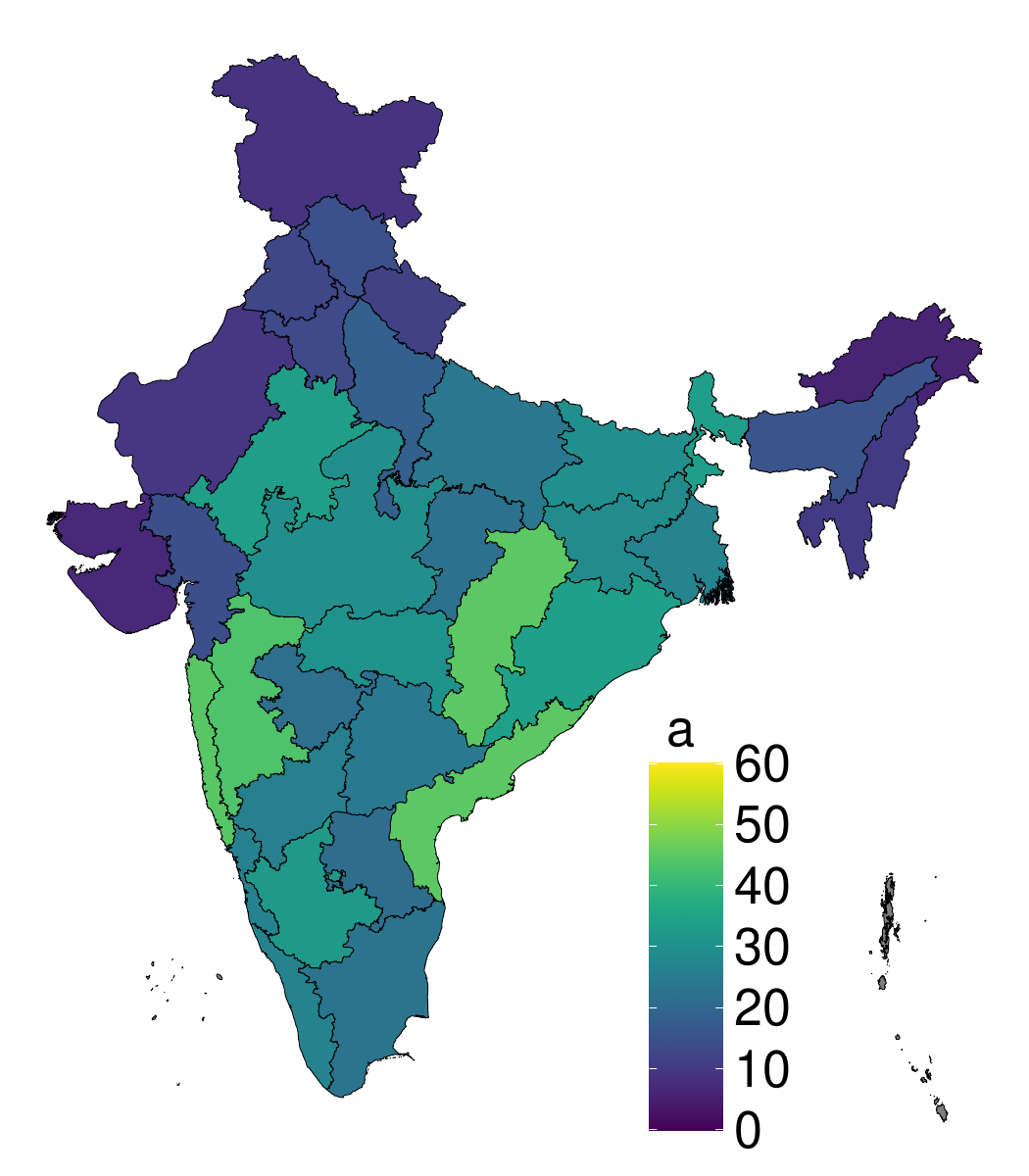}
    \end{subfigure}
    \hfill
    \begin{subfigure}[b]{0.3\textwidth}
        \centering
        \includegraphics[width=\textwidth]{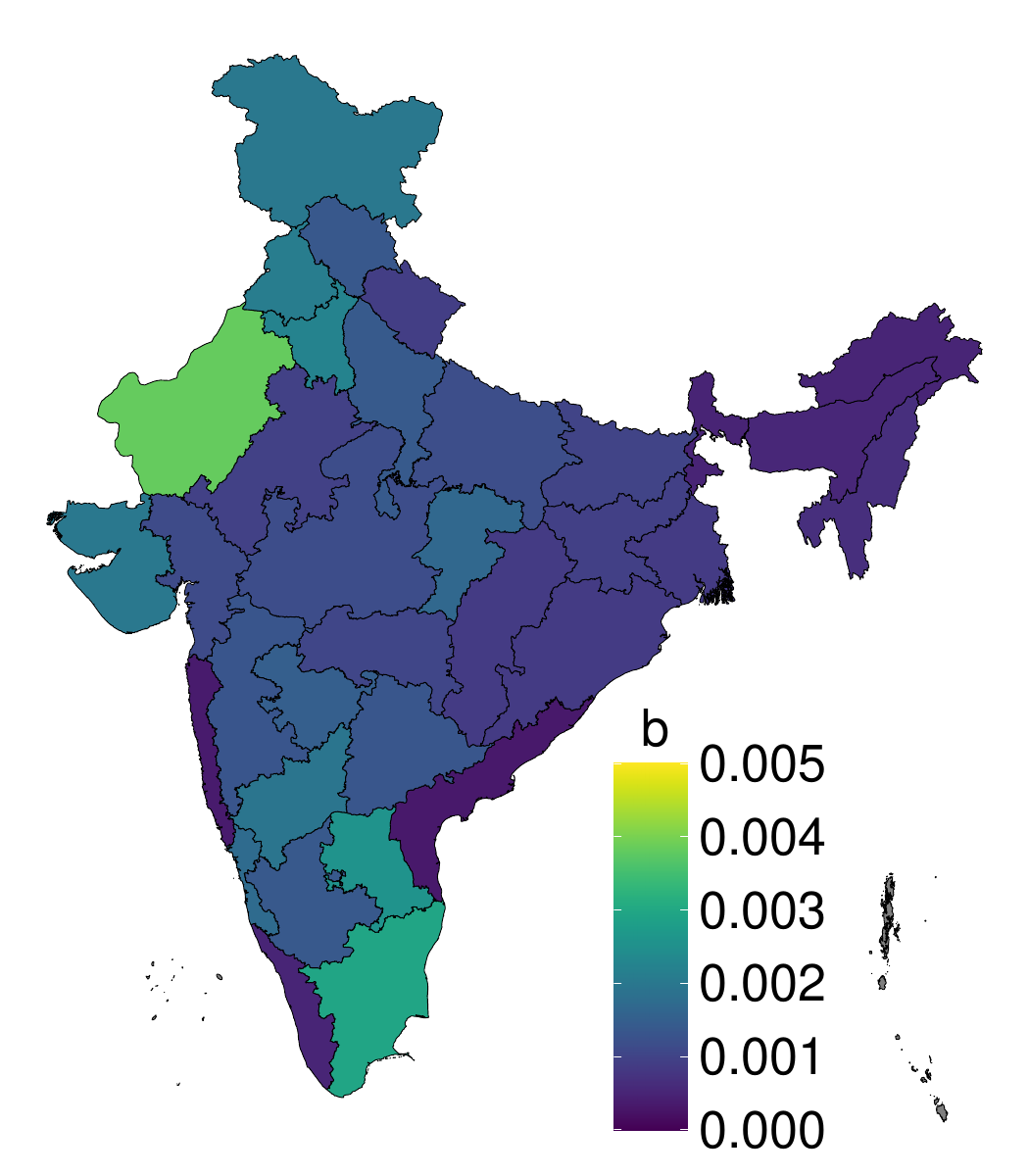}
    \end{subfigure}
    \hfill
    \begin{subfigure}[b]{0.3\textwidth}
        \centering
        \includegraphics[width=\textwidth]{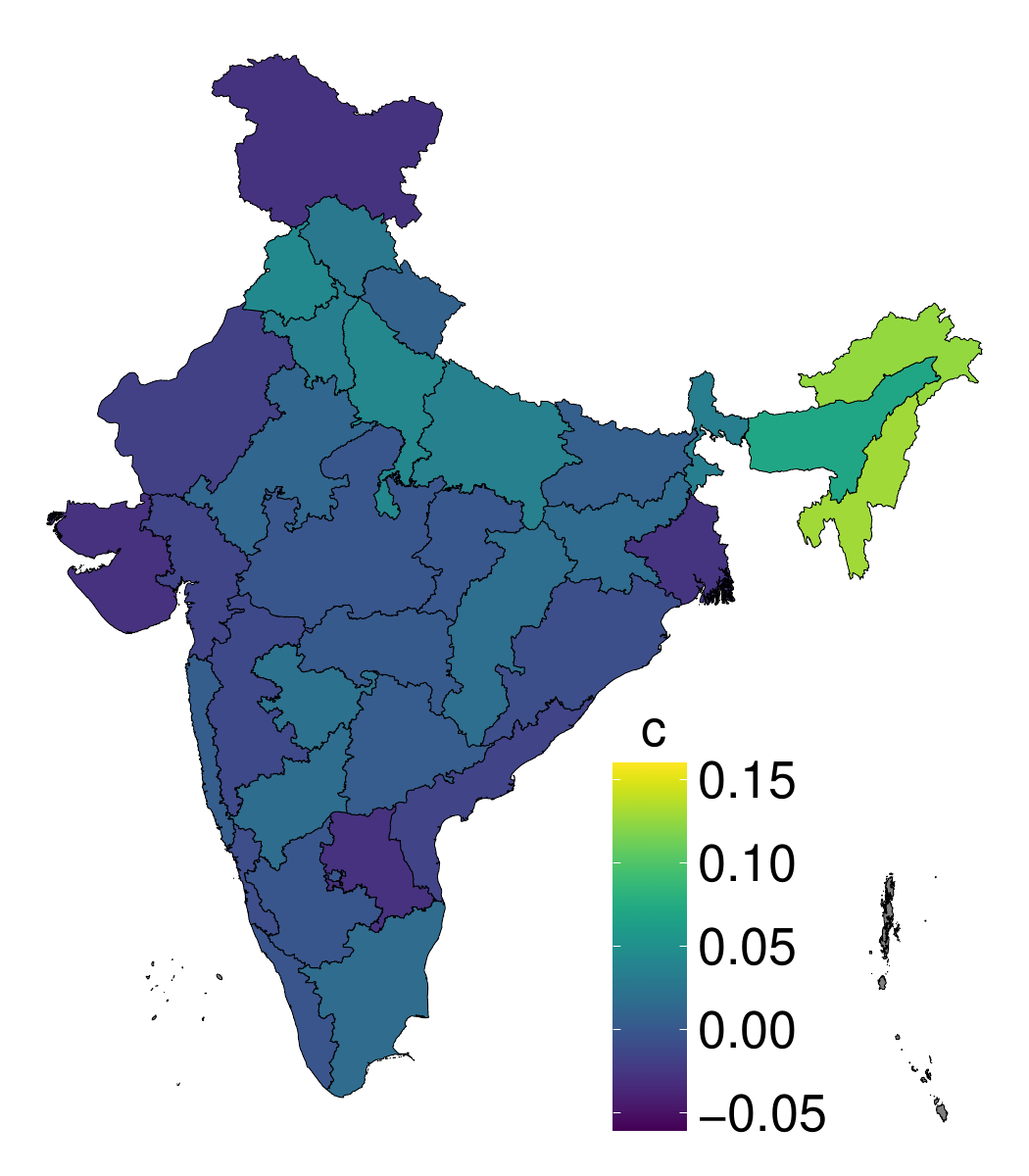}
    \end{subfigure}

    \vspace{1em}

    \begin{subfigure}[b]{0.3\textwidth}
        \centering
        \includegraphics[width=\textwidth]{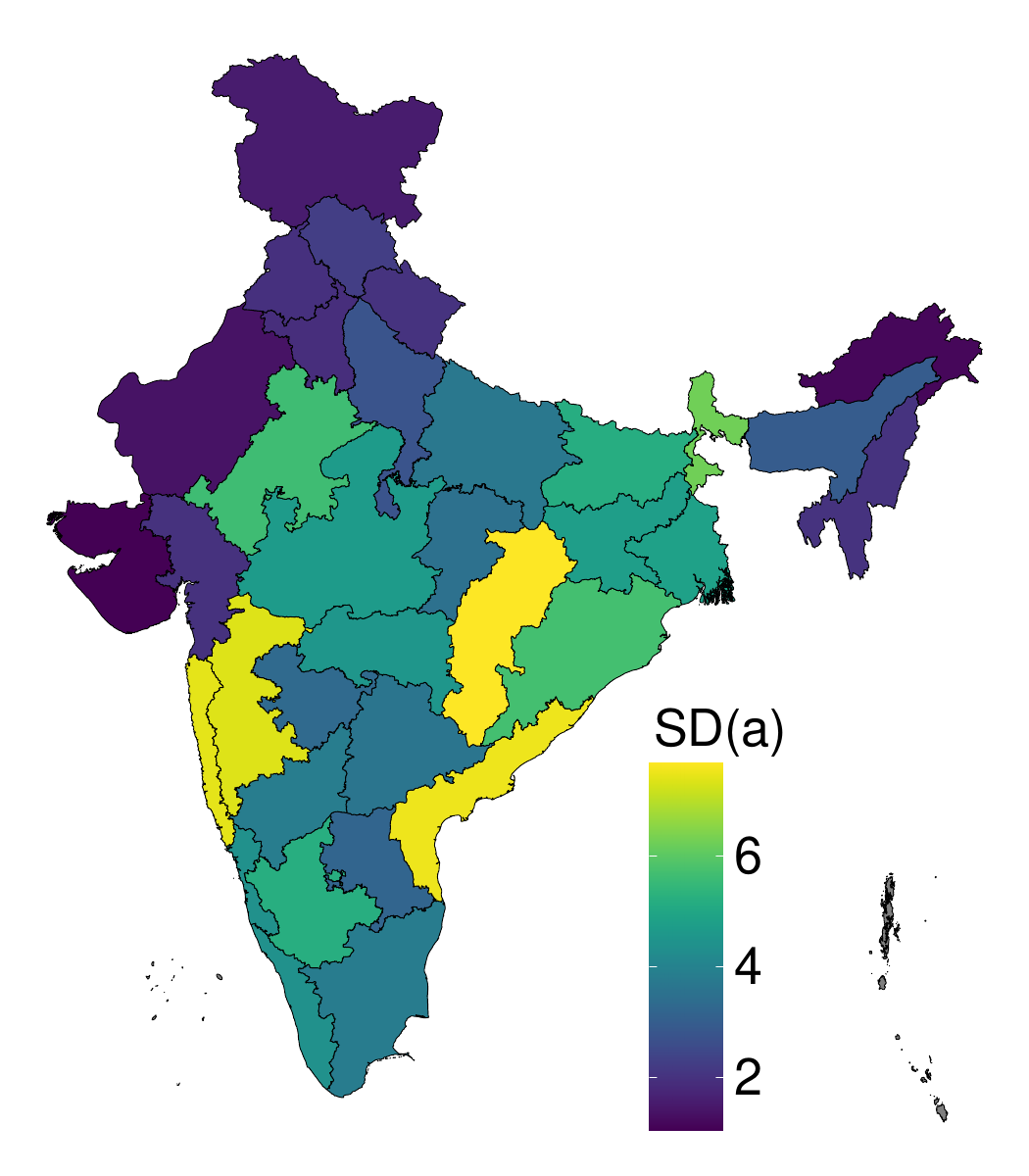}
    \end{subfigure}
    \hfill
    \begin{subfigure}[b]{0.3\textwidth}
        \centering
        \includegraphics[width=\textwidth]{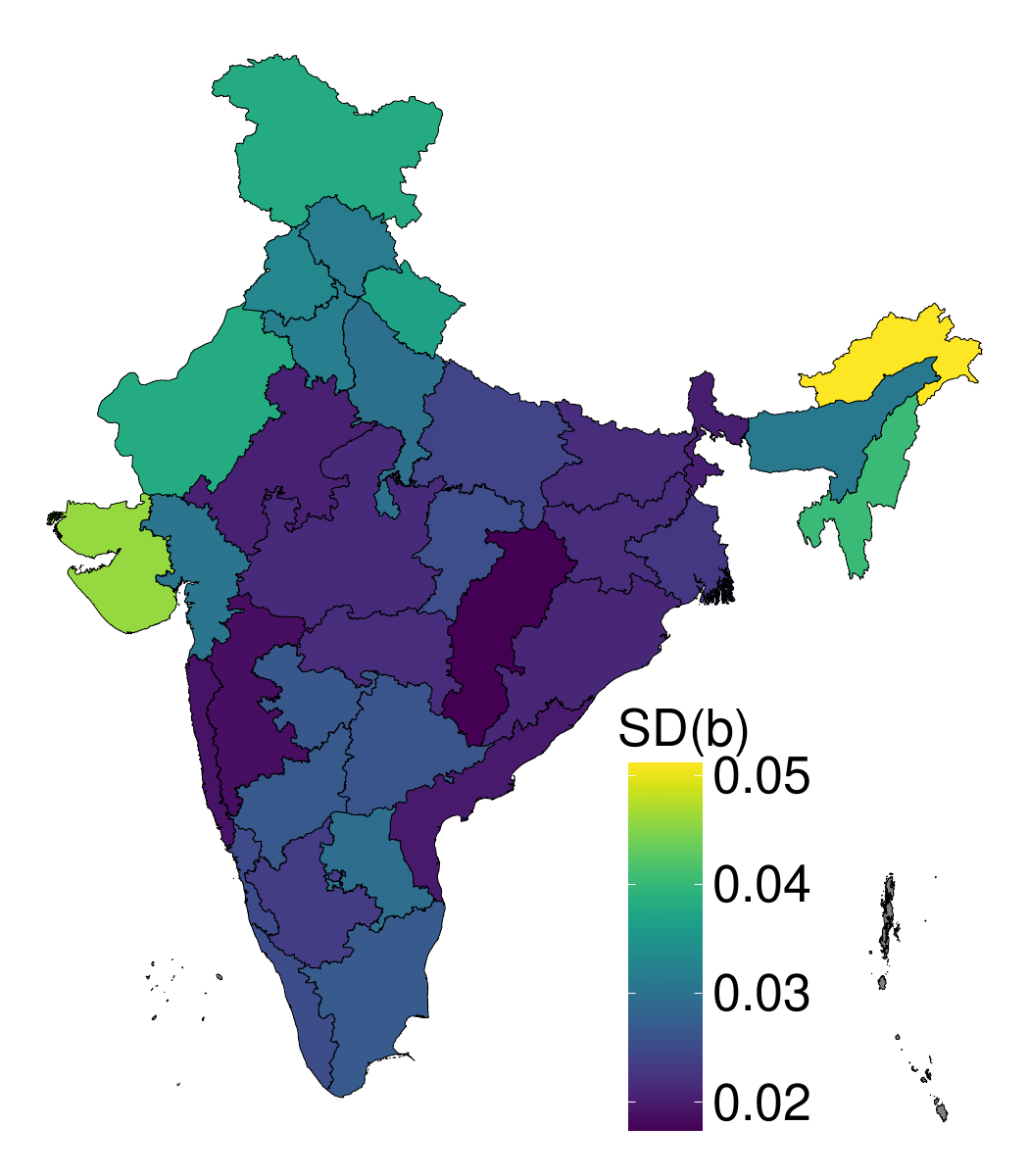}
    \end{subfigure}
    \hfill
    \begin{subfigure}[b]{0.3\textwidth}
        \centering
        \includegraphics[width=\textwidth]{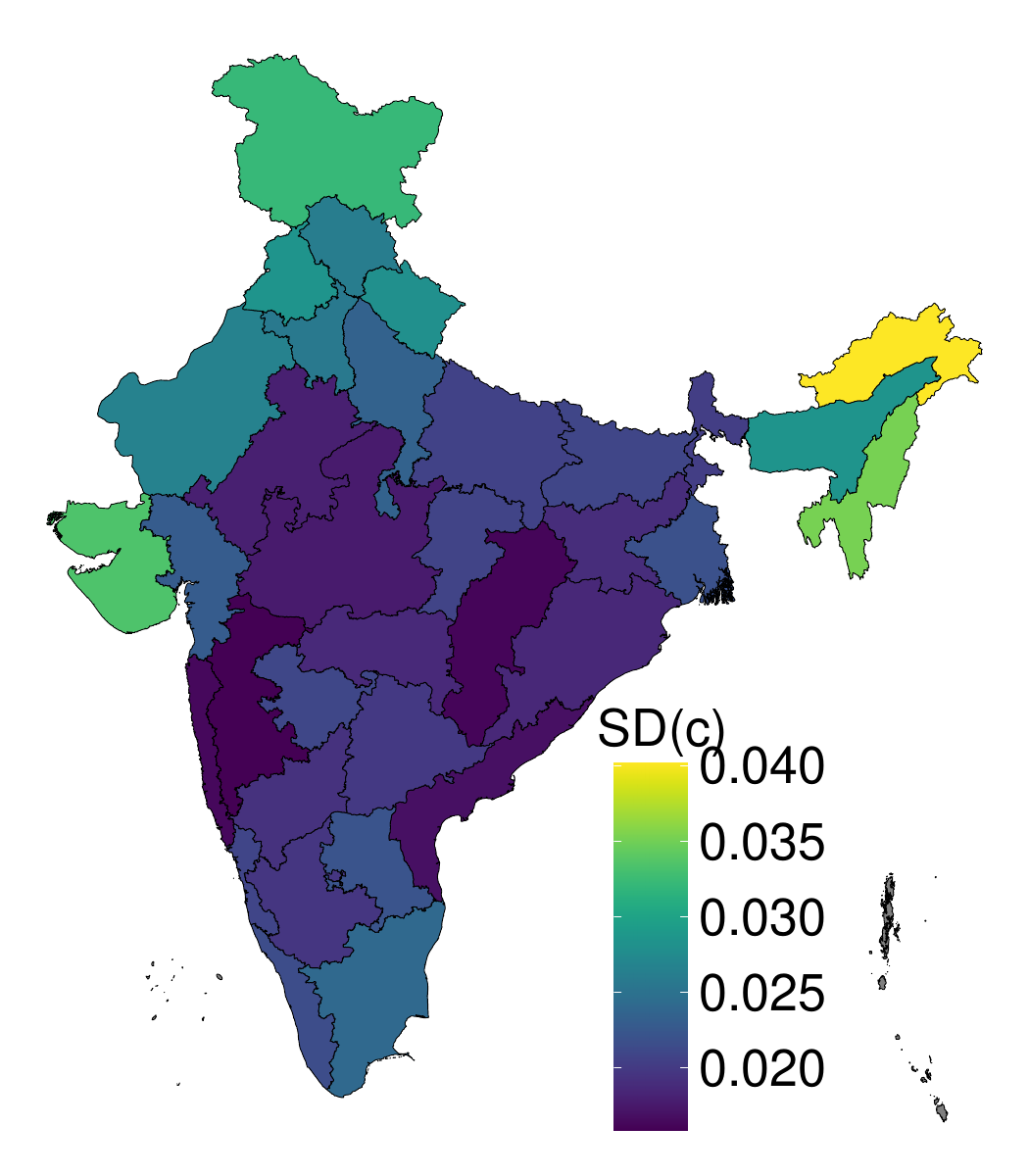}
    \end{subfigure}

    \caption{Posterior means (top) and SDs (bottom) of $a_i$s, $b_i$s, and $c_i$s based on fitting the CAR-ICAR model.}
    \label{fig:posmean_possd_realdata}
\end{figure}

We report the posterior means and SDs of $a_i$s, $b_i$s, and $c_i$s in Figure \ref{fig:posmean_possd_realdata}. Besides, the posterior mean and SD of $\rho$ are 0.9309 and 0.0114, respectively. The spatial patterns of the coefficients are similar to the ones observed in Figure \ref{fig:fig_mles_abc}, which indicates that the CAR-ICAR model can capture the spatial patterns of the coefficients correctly. The posterior means of $a_i$s, i.e., the shape parameters of the gamma model, are higher in the middle, eastern, and southern parts of India. At the same time, they are low in the northwestern parts of India and also near the northeastern meteorological subdivisions. Several eastern and western Indian subdivisions showcase higher posterior SD than others. The posterior means of $b_i$s, the intercept-related parameters of the gamma model, are higher in the northwestern and northeastern meteorological subdivisions. At the same time, they are lowest in the middle parts of the country. Finally, a positive $c_i$ indicates a negative trend follows from \eqref{eq:interpretation}. We observe positive estimates of $c_i$s throughout the country, indicating a negative trend in the amount of monsoon total rainfall. However, due to high posterior SDs, most of these estimates are insignificant except for three meteorological subdivisions: 1. Arunachal Pradesh, 2. Assam and Meghalaya, and 3. Nagaland, Manipur, Mizoram, and Tripura. A significantly negative trend for these Himalayan regions indicates a decline in mean total monsoon rainfall, and thus requires further environmental assessment.

\section{Concluding remarks}
\label{sec:conclusion}

Evaluating the availability of rainfall water is essential for rainfed agriculture, which is predominantly practiced in India. Considering potential trends in rainfall patterns due to climate change, we develop a statistical model to analyze monsoon total rainfall data for 34 meteorological subdivisions of mainland India from 1951 to 2014. Our approach models the marginal distributions using a gamma regression model while capturing dependence through a Gaussian conditional autoregressive (CAR) copula model. Given the natural variation in monsoon total rainfall across the diverse climatic regions of the country, we incorporate gamma regression with spatially varying coefficients (SVCs) within a latent Gaussian model framework. The dependence structure in both the likelihood and prior layers is governed by the neighborhood relationships of the regions, where we examine both CAR and intrinsic CAR structures for the priors. We employ Markov chain Monte Carlo algorithms to facilitate Bayesian inference, precisely a combination of block Metropolis-Hastings and single-component Metropolis-Hastings steps within a Gibbs sampler. Additionally, the proposed methodology effectively imputes missing data within the Gibbs sampling. Our model demonstrates superior performance in simulation studies compared to alternatives that do not account for dependence structures in the data or prior layers. Applying our methodology to the Indian areal rainfall dataset, we estimate model parameters, correctly quantify the uncertainties, and investigate the potential impact of climate change on rainfall patterns across India. The mean annual rainfall shows a significant negative trend for three meteorological subdivisions of northeastern India. 

A drawback of the proposed methodology is a specific dependence structure across space determined by the neighborhood structure of India's meteorological subdivisions. The literature contains a large number of copula models for Gaussian graphical models. However, our approach is parsimonious and implementable in all scenarios where a CAR model is applicable. The significance of our proposed methodology extends beyond rainfall modeling. It provides a generalizable framework for analyzing other spatially dependent climatic variables, such as temperature and humidity, which exhibit complex dependence structures. This approach is instrumental in climate impact assessments, where accurate predictions of future rainfall trends can inform adaptation and mitigation strategies \citep{fischer2013robust}. Incorporating SVCs into rainfall modeling is crucial for accurately capturing the inherent heterogeneity of precipitation patterns across different regions. Traditional models with fixed coefficients often fail to account for local variations, leading to oversimplified representations of rainfall processes. Allowing model parameters to change spatially enables SVC models to better represent the complex interactions between environmental factors and rainfall distribution. This enhanced flexibility improves the precision of hydrological predictions, which is vital for effective water resource management and flood risk assessment. For instance, studies have demonstrated that SVC models can effectively capture both short-range and long-range spatial dependencies in hydrological processes, leading to more accurate estimations of mean annual runoff coefficients \citep{bakar2020interpolation, gelfand2003spatial,arnaud2002influence}. Integrating SVCs into rainfall models enables a more nuanced understanding of spatial variability, resulting in more reliable and region-specific hydrological insights.

Our model specification incorporates spatial structure at two levels: in the marginal parameters through CAR/ICAR priors, and in the data layer through the CAR-copula dependence. This two-layer dependence is conceptually appealing, reflecting epistemic uncertainty in regional parameters and aleatoric uncertainty in rainfall totals. Similar hierarchical structures have been studied in spatial statistics \citep{reich2006effects, hughes2013dimension, hanks2015restricted}, where spatial random effects can coexist at multiple levels. While our results indicate that both layers contribute complementary information, potential confounding between them cannot be entirely ruled out. Approaches such as sum-to-zero constraints or restricted spatial regression \citep{reich2006effects, hughes2013dimension} could help formally orthogonalize the two components, and we view this as an important direction for future work. In this work, we have focused on a gamma regression framework with time as the primary covariate, but the approach naturally extends to a broader class of models. Specifically, any distribution from the exponential family can be used for the marginal distributions, with the mean parameter linked to multiple temporal covariates (such as year, global climate indices, or atmospheric variables) through a generalized linear model structure. Spatial variation can be incorporated by modeling regression coefficients as SVCs with appropriate spatial priors. Moreover, dependence across regions can be captured through a Gaussian copula with CAR or GMRF correlation structures, combined with marginal distributions drawn from the exponential family. This generalization enables more flexible spatio-temporal modeling of rainfall or other environmental responses within a unified framework.

\section*{Acknowledgement}
The authors thank an Associate Editor and four anonymous Reviewers for suggesting several crucial changes that substantially improved the manuscript. Additionally, the authors would like to thank Aritra Basak for providing the relevant codes for preliminary analysis and the adjacency matrix used in our paper. The research of the second author is supported by the Start-up Research Grant (SRG), Science and Engineering Research Board, Department of Science \& Technology, Government of India, with Award No. SERB-MATH-2023632.

\section*{Supplementary information}

Codes (written in \texttt{R}) for implementing our proposed methodology and the dataset analyzed in this paper are provided in the Supplementary Material. Additionally, the extension of gamma regression to a more general framework for non-Gaussian areal datasets, further exploratory analysis, univariate trace plots with effective sample sizes, and model diagnostics are also provided.

\section*{Conflict of interest statement}

No potential conflict of interest was reported by the authors.

\bibliography{references}

\end{document}